\documentclass[aps,prl,twocolumn,preprintnumbers,accepted=2024-12-11,nofootinbib,a4paper]{quantumarticle}
\pdfoutput=1
\usepackage{amsmath, mathrsfs, amssymb,amsfonts,amsthm,graphicx, epsf, dcolumn, yfonts, dsfont}
\usepackage[hyperfootnotes=true]{hyperref}
\usepackage[normalem]{ulem}
\usepackage{color}
\usepackage{slashed}
\usepackage{setspace}
\usepackage{cancel}
\usepackage{wasysym}
\usepackage[lofdepth,lotdepth]{subfig}
\usepackage{float}
\usepackage[utf8]{inputenc}
\usepackage{ragged2e}
\usepackage[numbers]{natbib}

 \newcommand{\be}{\begin{equation}}
 \newcommand{\ee}{\end{equation}}
 \newcommand{\bea}{\begin{eqnarray}}
 \newcommand{\eea}{\end{eqnarray}}

\def\del{\partial}

\def\cqstate{\varrho}

  \newtheorem{assumption}{Assumption}

\newcommand{\beq}{\begin{equation}}
\newcommand{\eeq}{\end{equation}}

\def\dxi{{\dot\xi}}

\def\dxi{ \xi}

\def\lapsh{[N,\vec{N}]}
\def\qmatterham{\hat{H}^{(m)}}

\renewcommand*{\thefootnote}{\fnsymbol{footnote}}

\newcommand{\ket}[1]{|#1\rangle}

\newcommand{\tr}{\mathop{\mathsf{tr}}\nolimits}
\begin{document}

\title{A healthier semi-classical dynamics}
\author{Isaac Layton}
\affiliation{Department of Physics and Astronomy, University College London, Gower Street, London WC1E 6BT, United Kingdom}
\author{Jonathan Oppenheim}
\affiliation{Department of Physics and Astronomy, University College London, Gower Street, London WC1E 6BT, United Kingdom}

\author{Zachary Weller-Davies}
\affiliation{Perimeter Institute for Theoretical Physics, Waterloo, Ontario, Canada}
\affiliation{Department of Physics and Astronomy, University College London, Gower Street, London WC1E 6BT, United Kingdom}
\date{}

\begin{abstract} \vspace{-10mm}
We study the back-reaction of quantum systems onto classical ones. Taking the starting point that semi-classical physics should be described at all times by a point in classical phase space and a quantum state in Hilbert space, we consider an unravelling approach, describing the system in terms of a classical-quantum trajectory. We derive the general form of the dynamics under the assumptions that the classical trajectories are continuous and the evolution is autonomous, and the requirement
that the dynamics is linear and completely positive in the combined classical-quantum state. This requirement is necessary in order to consistently describe probabilities, and forces the dynamics to be stochastic when the back-reaction is non-zero.  The resulting equations of motion have features reminiscent of the standard semi-classical equations of motion, but since the resulting dynamics is linear in the combined classical-quantum state, it does not lead to the pathologies which usually follow from evolution laws based on expectation values. In particular, the evolution laws we present account for correlations between the classical and quantum system, which resolves issues associated with other semi-classical approaches. In addition, despite a breakdown of predictability in the classical degrees of freedom, the quantum state evolves deterministically conditioned on the classical trajectory, provided a trade-off between decoherence and diffusion is saturated. As a result, the quantum state remains pure when conditioned on the classical trajectory. To illustrate these points, we numerically simulate a number of semi-classical toy models, including one of vacuum fluctuations as a source driving the expansion of the universe. Finally, we discuss the application of these results to semi-classical gravity, and the black-hole information problem.
\end{abstract}

\renewcommand*{\thefootnote}{\arabic{footnote}}
\setcounter{footnote}{0}

Many of the difficulties in modern physics, such as the correct description of black holes,  inflationary cosmology, or measurement, seem to occur in the semi-classical regime.  By this, we mean a regime in which classical and quantum systems appear to coexist and interact \cite{boucher1988semiclassical,diosi1995quantum} (rather than a regime in which the WKB approximation holds \cite{bender2013advanced,sakurai2020modern}). There are several reasons for choosing a semi-classical description: there may exist no fully quantum description, such as in the case of gravity; a full quantum theory exists, but is computationally unattainable; or that some fundamental degree of freedom, such as the measurement record of the experimenter or spacetime geometry, is presupposed to be classical in nature. Regardless of whether semi-classicality is viewed as effective or fundamental, it is important to understand which dynamics of classical and quantum systems are consistent, and which cause the semi-classical description to break down.

The history of defining consistent semi-classical dynamics, i.e. a consistent coupling between classical and quantum systems, has been one of controversy \cite{cecile2011role,witten1962gravitation,eppley1977necessity}. When the quantum system is controlled by the classical one without back-reaction,
the dynamics is described by unitary quantum mechanics, with the quantum state $|\psi\rangle$ at time $t$ determined by a Hamiltonian $H$ that depends on classical degrees of freedom $z$
\begin{equation} \label{eq: cControl}
\frac{d |\psi\rangle}{dt}=-i H(z) |\psi\rangle.
\end{equation}
Such dynamics are consistent with a semi-classical description, in the sense that the standard rules of quantum and classical mechanics may be applied without modification to each system independently.

However, defining consistent dynamics where the classical system is affected by the quantum one, i.e. experiences back-reaction, has proved more difficult. In the case of gravity, the standard approach to include backreaction is via the semi-classical Einstein equations, which source the Einstein tensor $G_{\mu \nu}$ by the expectation value of the stress energy tensor $T_{\mu \nu}$ \cite{sato1950attempt,moller1962theories, rosenfeld1963quantization} 
 \begin{equation}\label{eq: semiclassEinstein}
    G_{\mu \nu} =  8 \pi G \langle T_{\mu \nu} \rangle
\end{equation}
 (we use units where $\hbar=c=1$, and $G$ here is the gravitational constant).
Assuming the quantum degrees of freedom evolve according to quantum field theory in curved spacetime, Equation \eqref{eq: cControl}  with $z=(g,\pi,N,\bar{N})$ (i.e. the gravitational degrees of freedom) and Equation \eqref{eq: semiclassEinstein}, together provide the standard theory of semi-classical gravity \cite{HorowitzSC,Hu:2008rga}. 

\begin{figure}
    \includegraphics[width=8cm,trim={0cm 2.5cm 0.8cm 0},clip]{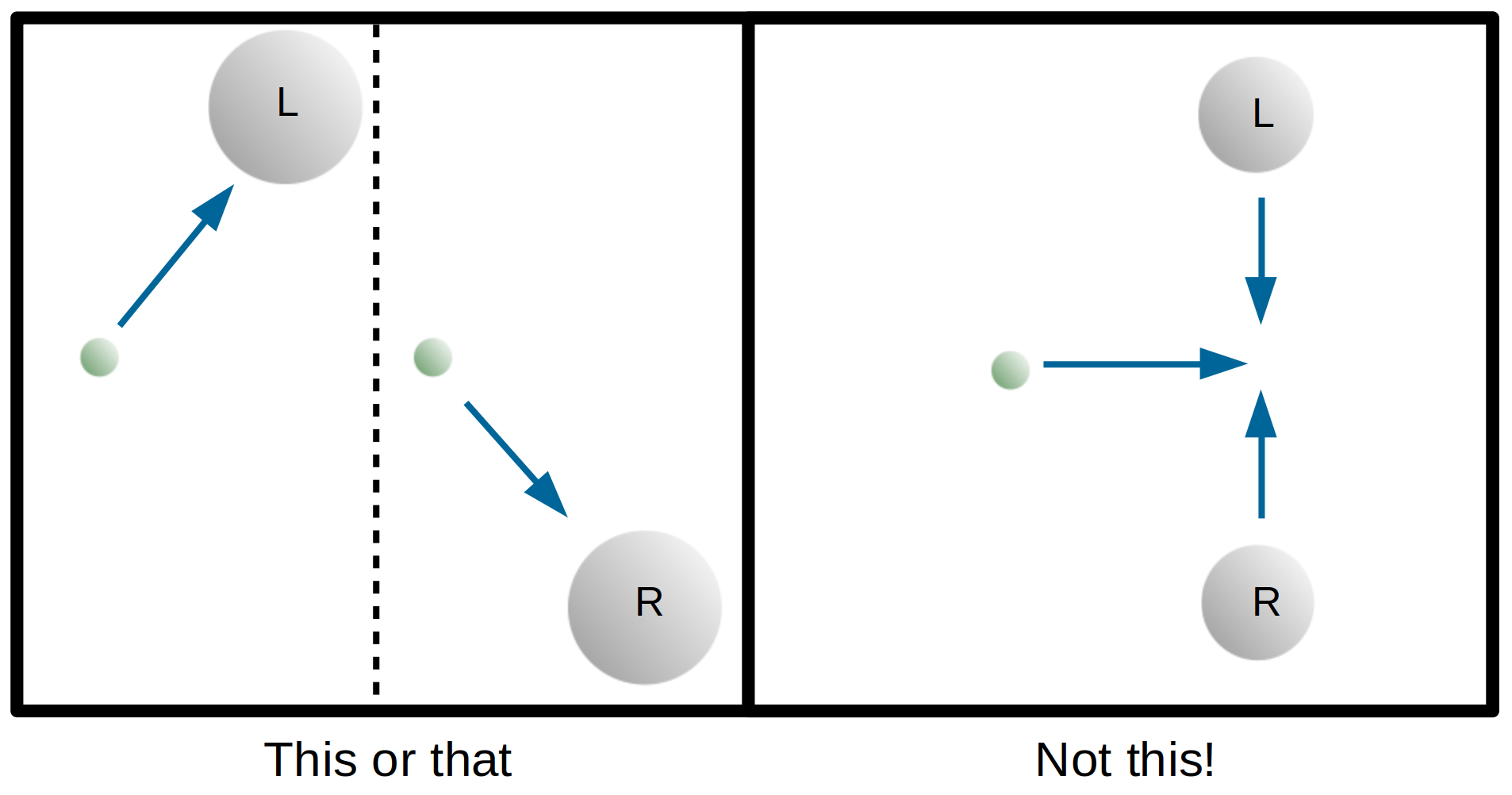}
    \caption{
    \footnotesize{A schematic showing the dynamics for both a theory that accounts for correlations (left) and the standard semi-classical theory (right). In both cases a test mass is released near a large mass that is described as a quantum system in the theory. The predictions are the same regardless of whether the large mass is in a superposition of spatial positions $|L\rangle$ and $|R\rangle$ or simply a classical mixture of the two; the standard semi-classical theory predicts the test mass will fall towards the average position of where the large mass might have been, while a semi-classical theory that accounts for correlations predicts that the test mass will either fall to the large mass being on the left or the large mass being on the right. Taking the expectation value of Einstein's equations destroys information about the correlation between the position of the large masses and the gravitational field. Similarly, a fully quantum theory that accounts for correlations predicts that the state of the gravitational field becomes entangled with the position of the planets \cite{Kabel:2022cje, Chen:2022wro}. }}\label{fig:notthismain}
\end{figure}

Although the equations of semi-classical gravity can be derived from effective low energy quantum gravity, they are commonly understood to fail when fluctuations of the stress-energy tensor are large in comparison to its mean value \cite{deWittQG, Kiefer:2004xyv, Veltman, Wallace:2021qyh,Jordanstability,Ford:1982wu,Kuo:1993if,wald1977back, Hu:2008rga,HartleHorowitz,boucher1988semiclassical}. However, the case where the fluctuations are significant are often precisely the regimes we most wish to understand, such as in considering the gravitational field associated to Schrodinger cat states of massive bodies \cite{page1981indirect,Arndt}, or vacuum fluctuations during inflation \cite{Hu:2000ns, Perez:2005gh, KieferPolarski,Kiefer:2006je}. In these cases, the equations of semiclassical gravity fail because they fail to allow for a build up of correlations between the classical and quantum degrees of freedom, as depicted on the right-hand side in Figure \ref{fig:notthismain}. For these regimes, background field methods are not appropriate, and an alternate effective theory of semi-classical gravity is required.

The possibility of an alternative approach to semi-classical physics than the one taken by semi-classical gravity has been considered for some time under the name of classical-quantum dynamics \cite{aleksandrov1981statistical,gerasimenko1982dynamical,boucher1988semiclassical,blanchard1993interaction,diosi1995quantum,DiosiHalliwel,HalliwelDH,2016Tilloy,tilloy2017principle,oppenheim_post-quantum_2018,CQPawula}. The earliest work in this direction attempted to derive a general form of consistent semi-classical dynamics from a series of postulates \cite{aleksandrov1981statistical,gerasimenko1982dynamical,boucher1988semiclassical}. However, in this case, the dynamics does not simultaneously satisfy two basic requirements for consistency, namely the positivity and linearity of the dynamics. Subsequently, pioneering later work by Blanchard and Jadczyk, and Diosi found examples of consistent linear and positive dynamics using the master equation formalism \cite{blanchard1993interaction,blanchard1995event,diosi1995quantum}. Recently, the general form of completely-positive classical-quantum master equations were characterised \cite{oppenheim_post-quantum_2018,CQPawula}, offering up the possibility of studying semi-classical dynamics in a general and consistent framework. 


In this paper we study unravellings \cite{risken1989fpe,Belavkin, Dalibard, Gardiner, Gisin:1989sx, GISIN19901, GisinMeas} of general classical-quantum master equations \cite{oppenheim_post-quantum_2018,CQPawula}. As we show, these provide a natural language to construct consistent semi-classical dynamics, with the dynamics explicitly taking the form of a stochastic modification to the standard semi-classical approach of semi-classical gravity. Agreeing with indications from earlier work, we find that consistent models of backreaction must be inherently stochastic, predicting probability distributions over classical trajectories in phase space and quantum trajectories in Hilbert space. Crucially, the additional stochasticity added to both the classical and quantum equations of motion is sufficient to restore the ability to correctly describe correlations between the classical and quantum sectors, as depicted on the left-hand side of Figure \ref{fig:notthismain}.

An early example of a classical-quantum unravelling was constructed in \cite{DiosiHalliwel} for the constant force case, using the formalism of continuous measurement theory \cite{2006Jacob,wiseman_milburn_2009}. In this paper we characterise the general form of classical-quantum unravellings, and do so axiomatically, requiring only that the dynamics be consistent, and assuming the dynamics to be autonomous and continuous in the classical degrees of freedom. This allows us to describe the general form of classical-quantum interactions in terms of classical-quantum trajectories. 

In addition to characterising the general form of classical-quantum unravellings, we are able to provide a number of key technical insights regarding their structure. Firstly, we are able to characterise for the first time the general conditions under which the dynamics are such that the quantum state conditioned on the classical trajectory remains pure. Secondly, we show that the consistent dynamics we provide can always be purified, such that quantum states remain pure conditioned on hidden degrees of freedom, by analogy with the church of the large Hilbert space of quantum mechanics. Finally, we demonstrate that classical-quantum dynamics completely parameterizes continuous measurement and (non-Markovian) feedback procedures, showing the equivalence of the continuous measurement \cite{kafri2014classical, diosi2014hybrid,2016Tilloy} and hybrid \cite{diosi2011gravity, oppenheim_post-quantum_2018, oppenheim_post-quantum_2018} approaches to continuous classical-quantum coupling, which have  often been treated as being mathematically distinct in discussions of classical-quantum gravity \cite{Gro_ardt_2022}.

By using these insights from the study of classical-quantum unravellings, we are able to provide a number of technical tools to study and improve on existing semi-classical methods. Our main contribution is to construct a general form of consistent Hamiltonian semi-classical dynamics, and find necessary and sufficient conditions for its positivity. In doing so, we are able to rigorously characterise the regime of validity of the standard semi-classical approach, such as that used by semi-classical gravity, as well as providing a regime that matches how semi-classical gravity is often used in practice. By showing that these typical uses arise in limits of the theory we present, we demonstrate how the stochastic dynamics we find provides a natural extension of semi-classical methods into regimes in which standard approaches breakdown.

Finally, we conclude by discussing how the formalism developed here may be applied to the study of semi-classical gravity, and what lessons the framework may provide for the study of problems such as the correct description of fluctuations in inflationary cosmology and the information paradox.

\section{Framework} \label{sec: overview}

In this section we set up the main formalism and discuss the important concepts for our present study of semi-classical dynamics. Further information on the formalism of classical-quantum states and observables may be found in \cite{aleksandrov1981statistical,gerasimenko1982dynamical},
while the formalism of stochastic trajectories in the context of continuous measurement theory may be found in \cite{2006Jacob,wiseman_milburn_2009}.

\subsection{Classical-quantum trajectories}

We start by defining the basic objects of a semi-classical description. Since the goal is to define the dynamics for a pair of systems, one classical and one quantum, we assume that at all times the semi-classical system is fully characterised by the pair $(z,\rho)$. Here $z$ denotes the classical degrees of freedom, i.e. a point in phase space $\mathcal{M}$, while $\rho$ denotes the quantum state i.e. a positive semi-definite operator on Hilbert space $\mathcal{H}$ with $\tr{\rho}=1$. The full evolution of the semi-classical system is thus characterised by the classical-quantum pair at each point in time $t$, which we denote by $(Z_t,\rho_t)$. That is, we assume the joint system should be described by a classical trajectory $\{Z_t\}_{t\geq0}$ in phase space, and a quantum trajectory $\{\rho_t\}_{t\geq0}$ in (the space of density operators on) Hilbert space. While a priori we do not assume that a semi-classical description should imply pure quantum states, if a dynamics starts in a pure quantum state, and preserves the purity of that state, then the corresponding normalised state vectors will be written as $|\psi\rangle_t$ and $\{|\psi\rangle_t\}_{t\geq0}$ denotes the corresponding trajectory in Hilbert space. 

\begin{figure}[]
    \centering
    \includegraphics[width=8.8cm,trim={0.8cm 2.4cm 2.4cm 2.2cm},clip]{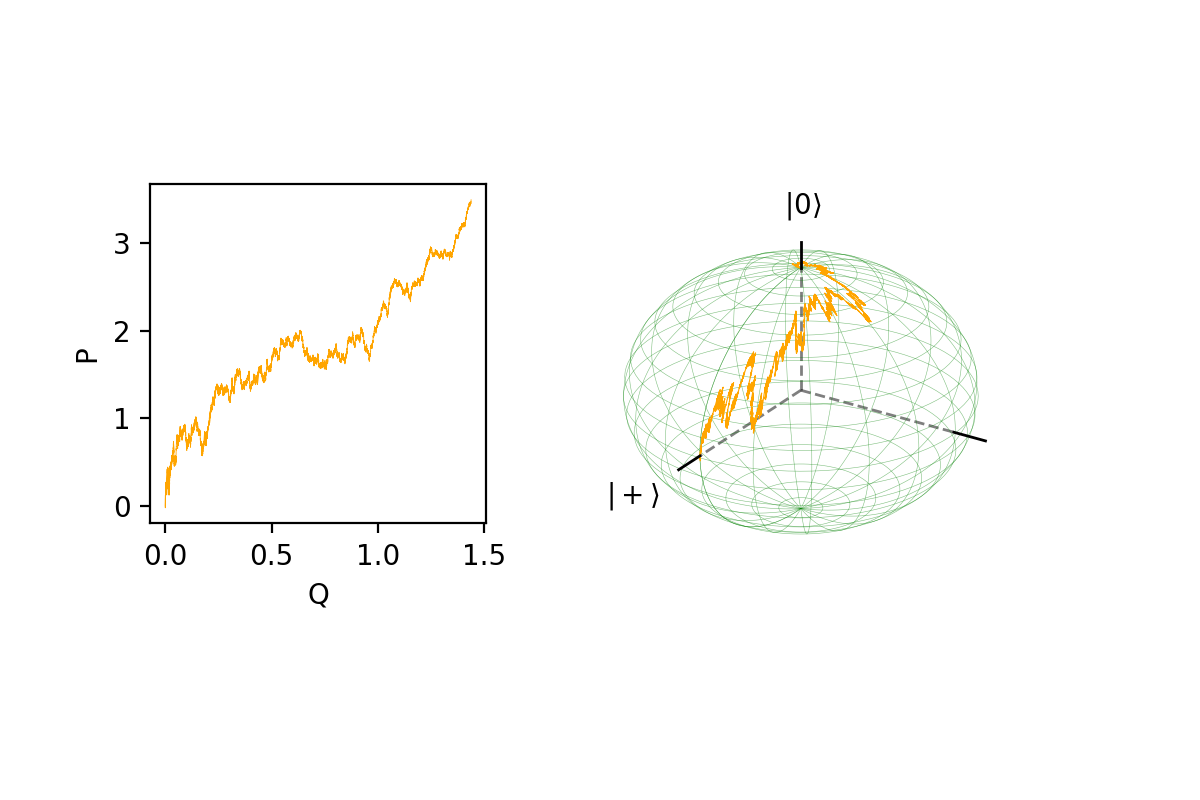}
    \caption{\footnotesize{A numerically simulated classical-quantum trajectory for a classical particle interacting with a qubit, represented by a classical trajectory in phase space (left) and a quantum trajectory on the Bloch sphere (right). The classical-quantum Hamiltonian is such that the classical system experiences a force either up or down depending on whether the state is $|0\rangle$ or $|1\rangle$ – the quantum state is then chosen to evolve starting in the superposition state $|+\rangle$. Initially starting at the origin in phase space, the classical system follows a stochastic trajectory with positive drift, agreeing with the evolution of the quantum state, which follows a path on the surface of the Bloch sphere before reaching the fixed point $|0\rangle$. The classical trajectory serves as a measurement record of the value of the qubit, and conditioned upon it the quantum state remains pure at all times. Since the classical particle's motion is stochastic, it takes some time to resolve the value of the qubit. The trajectories shown here should be contrasted with the standard semi-classical prediction, which predicts zero drift in momentum, and quantum evolution corresponding to a rotation about the $z$ axis of the Bloch sphere. Further details, such as the specific Hamiltonian studied and the initial conditions, may be found in Appendix \ref{sec: toy}. }}
    
    \label{fig:single_qubit_traj}
\end{figure}

As we shall see, standard approaches to semi-classical dynamics typically assume deterministic classical-quantum trajectories. However, more generally one could consider dynamics that generate probability distributions over classical trajectories in phase space and quantum trajectories in Hilbert space. Mathematically, this means treating ($Z_t,\rho_t)$ as a pair of coupled stochastic processes \cite{oskendal}. For what follows, it is sufficient to note that as stochastic processes, $Z_t$ and $\rho_t$ define random variables, taking values in phase space and states in Hilbert space respectively, at each time $t$. If the function taking $t$ to one of these random variables is continuous for every realisation of the random variables, we say that the stochastic process has continuous sample paths. Functions of $Z_t$ and $\rho_t$ become themselves random processes when evaluated at each time, and include real and operator valued functions such as the trace $\tr{\rho_t}$ or Hamiltonian $H(Z_t)$. The average value of a general function $f(Z_t,\rho_t)$ with respect to the probability measure is computed by the expectation $\mathbb{E}[f(Z_t,\rho_t)]$, and it will generically be the case that $\mathbb{E}[f(Z_t)g(\rho_t)]\neq\mathbb{E}[f(Z_t)]\mathbb{E}[g(\rho_t)]$ i.e. that the classical and quantum random variables are correlated. Note that this expectation is distinct from taking the trace of the quantum state with respect to an observable, $\tr{A \rho_t}$, which here computes a random variable describing the average outcome of a quantum measurement of the observable $A$ for each realisation of $\rho_t$.

\subsection{Observables and states}
Observables of the classical-quantum system are defined by the usual algebra of quantum observables with additional dependence on the classical degrees of freedom and are denoted $A(z), B(z),$ et cetera. Since both $Z_t$ and $\rho_t$ are random variables, it follows that the expectation value of a classical-quantum observable $\langle \langle A \rangle \rangle$ must be considered with respect to both the typical trace $\tr$ of a quantum state and the average over possible outcomes of these random variables i.e. 
\begin{equation} \label{eq: CQexp}
    \langle \langle A \rangle \rangle = \mathbb{E}[\tr\{A(Z_t)\rho_t\}],
\end{equation} where the double angled brackets here are intended to remind one that the average is taken both over the quantum state and as random variables. In the same way that $\tr\{A\rho \}$ in standard quantum theory predicts expectation values and probabilities for all possible quantum experiments by suitable choices of $A$, all outcomes of classical-quantum  experiments performed at time $t$ can be found by computing quantities of the form given in Eq. \eqref{eq: CQexp}. 

Although expectation values of classical-quantum observables are computable directly from $Z_t$ and $\rho_t$, the quantity that is sufficient to predict any outcome of a measurement at time $t$ is the \textit{classical-quantum state}
\begin{equation}\label{eq: CQstateDefinition}
\cqstate(z,t) =  \mathbb{E}[\delta(z-Z_t) \rho_t].
\end{equation} Since the quantum state $\rho_t$ is positive and normalised at all times, it follows that $\cqstate(z,t)$ is positive semi-definite at all points in phase space and normalised such that $\int dz \tr\{\varrho(z,t)\}=1$. To see that expectation values can also be computed using the classical-quantum state, we note simply that note that the product of $\varrho(z,t)$ and an observable $A(z)$ may be integrated over phase space and traced over Hilbert space to compute the expectation value \begin{equation} \label{eq: CQexp2}
    \langle \langle A \rangle \rangle =\int dz \tr\{A(z)\varrho(z,t)\},
\end{equation} which is equivalent to the definition given in Equation \eqref{eq: CQexp}.

The classical-quantum state is the natural generalisation of the classical probability distribution and quantum density operator to the classical-quantum setting. When the classical degrees of freedom are traced over, $\rho(t)=\int dz \varrho(z,t)$ describes a density operator with no conditioning on the classical degrees of freedom. When the Hilbert space is traced over, $P(z,t)=\tr\{\varrho(z,t)\} $ describes the classical probability distribution in phase space without conditioning on a quantum measurement outcome. The full object encodes correlations between the two systems – if the classical and quantum systems are correlated as random variables, then $\mathbb{E}[\delta(z-Z_t) \rho_t]\neq \mathbb{E}[\delta(z-Z_t)]\mathbb{E}[\rho_t]$, and thus the classical-quantum state does not factorise i.e. $\varrho(z,t)\neq \rho(t) P(z,t)$. In this case, one may understand $\varrho(z,t)$ is the density operator one assigns to the quantum system given that the classical system is at $z\in\mathcal{M}$ at time t, multiplied by the probability of finding a classical system at $z$ in the first place.


\subsection{Consistent classical-quantum dynamics} \label{sec: consistent}

A key feature common to both classical probability theory and quantum theory is that the states in the theory, i.e. probability distributions $P(z,t)$ and density operators $\rho(t)$, are assumed to have positive and linear evolution. In this work we shall assume the same must hold for the classical-quantum state $\varrho(z,t)$, and we will take this to be required for any classical-quantum dynamics to be consistent. Since $\varrho(z,t)$ is defined as $\mathbb{E}[\delta(z-Z_t) \rho_t]$, requiring the consistency of the evolution of $\varrho(z,t)$ puts constraints on the possible allowed dynamics of $\rho_t$ and $Z_t$.

To see why positivity and linearity of the dynamics of $\varrho(z,t)$ are important properties for consistency, we first note that probability distributions, density operators, and classical-quantum states, all satisfy the requirements of a more general definition of \textit{states}. Specifically, statistical mixtures of different configurations are represented by convex combinations of states, and expectation values are computed by a map that is linear on both the states and observables of the theory. This means that the classical-quantum states we present here are the same as the states discussed in a more general class of probability theories known as GPTs \cite{barrett2007information,janotta2014generalized,galley2021nogo}. 

In such theories, the requirement for a consistent probabilistic interpretation of measurement outcomes leads to constraints on the evolution of states. If the evolution laws do not preserve the positivity of the state, then the theory will predict negative probabilities for some measurement outcomes. If the evolution laws of states are not linear, it leads to the non-physical result that outcomes of experiments depend on whether independent measurement records are deleted before or after the evolution. To see this, take $\omega_1,\omega_2$ to be two states of a theory, generated with probabilities $p$ and $(1-p)$ respectively, and define the evolution map $\mathcal{E}_t$ as the map which takes every initial state to its final state at time $t$. If an observer forgets which state was prepared immediately, they predict the final state to be $\mathcal{E}_t(p\omega_1 + (1-p) \omega_2)$, which should be the same as $p\mathcal{E}_t(\omega_1) + (1-p) \mathcal{E}_t(\omega_2)$, the state they predict if they wait to delete the measurement record until after the evolution \cite{Gisin:1989sx,barrett2007information,diosi2011short,oppenheim_post-quantum_2018}.

Although we have simply stated that the dynamics of $\varrho(z,t)$ should be positive, in the context of quantum mechanics, a stronger notion of positivity is required for the dynamics to be consistent when acting on part of a larger system. This is called \textit{complete-positivity}, and ensures that when the dynamics is applied to half of an entangled quantum state, negative probabilities do not arise \cite{nielsen2002quantum}. We will assume the same holds in the classical-quantum case.

Taken together, we thus will state that a dynamics is \textit{consistent with a classical-quantum description}, or simply \textit{consistent}, whenever the dynamics of $Z_t$ and $\rho_t$ lead to dynamics that is completely-positive and linear on the classical-quantum state $\varrho(z,t)$.  It is worth emphasising that this is a very weak set of conditions, and indeed is a strict subset of the consistency conditions of \cite{boucher1988semiclassical}. This being said, given that it was shown in the same work that no dynamics could be constructed that satisfies the full set of these consistency conditions, we see that taking a weaker set of conditions is a necessary step to constructing examples of consistent classical-quantum dynamics. 

Finally, it is important to emphasise that a consistent theory may still have primary quantities that uniquely determine the state of a system, which nevertheless do not themselves evolve linearly. For example, in the Madelung formulation of unitary quantum mechanics, an auxiliary  scalar and vector field are used to describe the quantum wavefunction, and evolve according to a set of non-linear equations \cite{madelung1927quantum,bohm1954model}. The distinction to be drawn is that objects such as these do not define states in the sense we have so far considered, since, for example, they are not related to the expectation values of the theory via a linear map. To check the consistency of dynamics in such theories, it is necessary to directly compute the evolution of the states of the theory. In what follows, this has an important consequence: the evolution equations of $Z_t$ and $\rho_t$ need not necessarily be linear, provided the overall dynamics induced on the classical-quantum state $\varrho(z,t)$ is.

\subsection{Conditional quantum states} \label{subsec: conditioning}
When a projective measurement is made in conventional quantum mechanics, the state of the quantum system conditioned on the measurement outcome is pure. However, if the measurement outcome is not conditioned on, the act of measurement will generically cause the quantum state to lose purity i.e. decohere. 

To see how this feature also arises in  classical-quantum systems, we first note that when an observer has some partial information $Y$, the conditional expectation of $f(Z_t,\rho_t)$, denoted $\mathbb{E}[f(Z_t,\rho_t)|
Y]$, defines the improved estimate based on this information. This information will typically correspond to knowing some classical quantity $y(Z_t)$ for a subset of time $t\in I$, and so is formally described by the $\sigma$-algebra generated by this set of random variables, which we denote $\sigma\{y(Z_s)\}_{s\in I}$ \cite{oskendal}.  

If an observer were to repeat an experiment many times, each time measuring the quantum system at time $t$ and recording no information about the classical system, the quantum state they would infer from their observations would be $\rho(t)=\mathbb{E}[\rho_t]$. However, if each time they performed the experiment they additionally recorded some partial information about the classical trajectory $Y=\sigma\{y(Z_s)\}_{s\in I}$, the quantum state determined from their observations would be the \textit{conditional quantum state} \cite{Leifer_2013}
\begin{equation}
    \rho(t|Y)=\mathbb{E}[\rho_t|Y].
\end{equation}
i.e. an average only over realisations of the quantum state $\rho_t$ that occur with the classical observation $Y$. It is straightforward to prove, using the Jensen inequality \cite{oskendal}, that the entropy $S$ of the unconditioned state will always be greater than or equal to the average entropy of the conditioned state i.e.
\begin{equation}
    S(\rho(t))\geq \mathbb{E}[S(\rho(t|Y))].
\end{equation} The above inequality tells us that if there are correlations between the classical and quantum systems i.e. $\mathbb{E}[f(Z_t)g(\rho_t)]\neq\mathbb{E}[f(Z_t)]\mathbb{E}[g(\rho_t)]$, then conditioning on a particular outcome of the classical trajectory will generically give us an improved estimate of the state of the quantum system. This improved estimation will correspond to a state with greater purity.

An important example of a conditioned quantum state is the quantum state conditioned on the full classical trajectory up to time $t$, $\rho(t|\sigma\{Z_s\}_{s\leq t})$. The information corresponding to this state is maximal, in the sense that an observer that has access to the full classical trajectory has as much information about the combined classical-quantum system as is possible without disturbing it with a quantum measurement. Perhaps surprisingly, even in this case, there can exist classical-quantum trajectories for which $\rho(t|\sigma\{Z_s\}_{s\leq t})\neq \rho_t$. When this is the case, the individual trajectories of $\rho_t$ are not physical. This is because in this case the entropy of the conditioned quantum state $\rho(t|Y)$ is greater than that of $\rho_t$, and thus the non-uniqueness of the decomposition of mixed quantum states means there are physically equivalent $\rho_t^\prime$ which cannot be experimentally distinguished. This is an exact analogue of the case considered in the unravelling of quantum master equations, where individual realisations of pure states of the quantum system only take on physical meaning when they are correlated with a measurement apparatus monitoring the system \cite{wiseman_milburn_2009}.

For this reason, we make a specific requirement on the representation of the classical-quantum dynamics we will study. Specifically, we will always choose to represent dynamics such that
\begin{equation}
    \rho_t=\rho(t|\sigma\{Z_s\}_{s\leq t}),
\end{equation} where this equality is taken to hold except on a set of measure zero i.e. almost surely. In more technical language, this means restricting to stochastic processes of the quantum state $\rho_t$ such that the they are adapted to the $\sigma$-algebra $\sigma\{Z_s\}_{s\leq t}$ \cite{oskendal}. To achieve this, it is necessary to allow the quantum system to be decomposed in terms of mixed states, $\rho_t$, rather than pure states $|\psi\rangle_t$. In particular, if one was restricted to decompose the quantum system in terms of $|\psi\rangle_t$, one would find that in any setting where the quantum system undergoes decoherence from the perspective of an observer with maximum information, one would be forced to decompose the dynamics into a particular non-unique choice of pure quantum states. By contrast, since classical probability distributions can always be uniquely decomposed in terms of delta functions $\delta(z-Z_t)$ in phase space, we suffer no loss of generality by assuming an unravelling of the classical system in terms of pure states i.e. in terms of $Z_t$.

\section{Standard semi-classical dynamics} \label{sec: standard}

Having defined the necessary formalism for studying classical-quantum systems via their trajectories, we now turn to consider their dynamics. In this section, we introduce the standard approach to constructing semi-classical dynamics, and study why it does not lead to a consistent treatment of semi-classical physics.

As pointed out in earlier work \cite{boucher1988semiclassical}, the equations of semi-classical gravity, Equation \eqref{eq: cControl} with $z$ representing the time local gravitational degrees of freedom and Equation \eqref{eq: semiclassEinstein}, may be understood as a special case of a more general approach taken to describe back-reaction. In this work we shall refer to this as the \textit{standard semi-classical approach}, though it is also known in other fields as mean field dynamics \cite{prezhdo1999mean} or Ehrenfest dynamics \cite{tully1990molecular}. Although distinct, the standard semi-classical approach also shares some similarities with the Born-Oppenheimer approximation in it does not allow correlations to build up between the two systems. 

To write down the standard semi-classical dynamics, we first assume the existence of a Hermitian operator-valued function of phase space $H(z)$, that we refer to as the classical-quantum Hamiltonian. This Hamiltonian describes both the standard classical and quantum Hamiltonians $H_C(z)$ and $H_Q$ respectively, as well as their interactions via a Hermitian operator-valued function of phase space $H_I(z)$, which arise in the following canonical decomposition of the classical-quantum Hamiltonian
\begin{equation} \label{eq: CQ_Hamiltonian}
    H(z)=H_C(z) \mathds{1}+ H_Q + H_I(z),
\end{equation} where $\mathds{1}$ is the identity operator on the Hilbert space.

For a given classical-quantum Hamiltonian $H(z)$, we write down the standard semi-classical dynamics as follows. The classical evolution is deterministic and has back-reaction given by the expectation value of the quantum state,
\begin{equation} \label{eq: hamSemiClass}
\begin{split} 
    & dZ_{t,i} =  \langle \{Z_{i}, H \} \rangle dt,
\end{split}
\end{equation}
where here $\{ \cdot ,\cdot \}$ denotes the Poisson bracket of classical mechanics and the angled brackets denote the inner product with respect to $|\psi\rangle_t$ i.e. $\langle A \rangle = \langle \psi | A | \psi \rangle $. Meanwhile, the quantum evolution is given by Hamiltonian evolution that depends on the phase space degree of freedom
\begin{equation}\label{eq: hamSemiQuant}
    d |\psi \rangle_t = -iH(Z_t)|\psi\rangle_t dt.
\end{equation} 
Note that this dynamics allows initial correlations between the classical and quantum sectors, and we ignore more pathological versions e.g. in which the phase-space dependence in Equation \eqref{eq: hamSemiQuant} is an ensemble average over the classical degrees of freedom.

There are number of features of the standard semi-classical equations, \eqref{eq: hamSemiClass} and \eqref{eq: hamSemiQuant}, that make the dynamics at least at first glance desirable. Firstly, when the interaction Hamiltonian, $H_I(z)$, is zero, it is straightforward to check that that the dynamics reduces to the standard equations of classical and quantum Hamiltonian mechanics. Secondly, even when the interaction Hamiltonian is non-zero, the dynamics retains a number of expected features: the classical system evolves continuously in phase space, the quantum state $\psi\rangle_t$ remains normalised, and the evolution laws are autonomous (i.e. only depend on $Z_s$ and $|\psi\rangle_s$ at $s=t$, with the coefficients having no explicit dependence on $t$) as is the case for standard classical or quantum mechanics.

However, it is well-known that the standard semi-classical dynamics comprising Equations \eqref{eq: hamSemiClass} and \eqref{eq: hamSemiQuant} 
lead to a number of violations of the standard principles of quantum theory, inducing a break-down of either operational no-signalling, the Born rule, or composition of quantum systems under the tensor product \cite{ eppley1977necessity, page1981indirect,2016Tilloy, galley2021nogo}. These arguments appeal to the fact that the dependence of $Z_t$ on $|\psi\rangle_t$, generated by Equation $\eqref{eq: hamSemiClass}$, means that the evolution of \eqref{eq: hamSemiQuant} need not even be a linear map \cite{1984DiosiSchrodingerNewton, alicki1995comment,vstelmachovivc2001dynamics}, and thus can be ruled due to the inconsistency of non-linear modifications to quantum mechanics \cite{GISIN19901,polchinski1991weinberg,abrams1998nonlinear,Kent2005}.

Alternatively, one can directly rule out the standard semi-classical dynamics as a consistent theory by observing that the evolution law for $\varrho(z,t)$ is non-linear. To see this, we first note that by use of the chain rule, 
\begin{equation} \label{eq: nonlinearME}
    d\varrho=\mathbb{E}[d \delta (z-Z_t) \rho_t]+\mathbb{E}[ \delta (z-Z_t) d\rho_t],
\end{equation} where here $\rho_t=|\psi\rangle_t \langle \psi|_t $. Using the chain rule with equation \eqref{eq: hamSemiClass} and \eqref{eq: hamSemiQuant} to compute both $d \delta (z-Z_t)$ and $d\rho_t$ to first order in $dt$, we find that
\begin{equation}
\begin{split}
    \frac{\partial \varrho}{\partial t}=&-\frac{\partial}{\partial z_i} \mathbb{E}[\tr\{ \{Z_i,H\} \rho_t \} \delta(z-Z_t) \rho_t ] \\
    &-i[H(z),\varrho],
\end{split}
\end{equation} where we assume summation over the repeated $i$ index. Since the first term contains two occurrences of $\rho_t$, it cannot be written as a linear equation in terms of $\varrho$, unless each $\{z_i,H\}$ were proportional to the identity i.e. unless the interaction Hamiltonian $H_I(z)$ were everywhere zero. We thus see that when there is quantum back-reaction on the classical system, the standard semi-classical dynamics induce an evolution on $\varrho$ that is necessarily non-linear, and therefore inconsistent, as discussed in section \ref{sec: consistent}.

\section{Main results} \label{sec: main}

Although the standard semi-classical approach satisfies a number of important properties, the resulting dynamics is not linear on the classical-quantum state $\varrho(z,t)$. In this section, we introduce linearity of this quantity as an explicit assumption, which we use to arrive at a general form of consistent evolution on combined classical-quantum trajectories $\{(Z_t, \rho_t)\}_{t\geq 0}$. This result is given by a pair of coupled stochastic differential equations, \eqref{eq: unravelingCQClass} and \eqref{eq: unravelingCQCQuantum}, and we refer the reader to \cite{oskendal,2006Jacob} for more details on this formalism.

We begin by making two basic assumptions guaranteeing the consistency of our dynamics:
\begin{assumption} \label{ass: trajectories}
Solutions to the dynamics are described by a probability distribution over classical-quantum trajectories i.e. $\{(Z_t, \rho_t)\}_{t\geq 0}$.
\end{assumption}

\begin{assumption}\label{ass: linear} The dynamics induces an evolution on $\varrho(z,t)=\mathbb{E}[\delta(z-Z_t)\rho_t]$ that is  completely-positive and linear.
\end{assumption}

\noindent To further constrain the possible form of dynamics, we also make two assumptions consistent with the features of the standard semi-classical dynamics:

\begin{assumption}\label{ass: markovian}
The dynamics is autonomous on the combined classical-quantum system, meaning that the evolution laws comprise (stochastic) differential equations of the variables $Z_t$ and $\rho_t$ that are autonomous with respect to time. 
\end{assumption}

\noindent Such dynamics is sometimes called time-independent Markovian classical-quantum dynamics, and we refer the reader to Appendix \ref{sec: MarkovianApp} for further discussion on classical-quantum time-(in)dependent Markovianity and time-local dynamics. Both time-dependent Markovian dynamics, and dynamics which depends on the entire classical trajectory $\{Z_s\}_{s\leq t}$, are encompassed by Assumption \ref{ass: markovian} since one can always embed a clock system and or a system with memory of the classical trajectory as auxiliary classical systems. 

\begin{assumption}\label{ass: continous}
The classical trajectories $\{Z_t\}_{t\geq0}$ have (almost surely) continuous sample paths.
\end{assumption}

To derive a general form of dynamics from these assumptions, we start by returning to the auxillary object $\cqstate(z,t)=\mathbb{E}[\delta(z-Z_t) \rho_t]$ . By Assumption \ref{ass: trajectories}, $\rho_t$ is positive semi-definite and $\tr \rho_t =1$, and thus describes a classical-quantum state, as defined in Equation \eqref{eq: CQstateDefinition}.
 
The dynamics generating $\{Z_t\}_{t>0}$ and $\{\rho_t\}_{t>0}$ induce a dynamics on $\cqstate(z,t)$. By Assumption \ref{ass: linear} the dynamics on the state $\cqstate(z,t)$ must be linear and completely positive. Moreover, since at each time $\cqstate(z,t)$ is a classical-quantum state, the dynamics must also be norm preserving. Such dynamics has recently been characterized in a classical-quantum version of Kraus' theorem \cite{poulinPC,oppenheim_post-quantum_2018,CQPawula}. Furthermore, by studying the conditions of complete positivity, it has been shown \cite{CQPawula} that when the dynamics are autonomous (Assumption \ref{ass: markovian}), there is a unique family of CQ master equations with continuous trajectories in phase space (Assumption \ref{ass: continous}). 

For a given set of $p$ traceless and orthogonal operators $L_\alpha$ defined on the Hilbert space, and $n$ degrees of freedom $z_i$ in the phase space, these master equations are specified by a complex positive semi-definite $p\times p$ matrix $D_0$, a complex $n\times p$ matrix $D_1$, a real positive semi-definite $n\times n$ matrix $D_2$, a real vector of length $n$, $D_1^C$, and a Hermitian quantum operator $H$. All of the $D$ matrices and $H$ may themselves be functions of phase space. The general form of completely positive dynamics is then
\begin{equation}\label{eq: continuousME}
\begin{split}
\frac{\partial \cqstate(z,t)}{\partial t} & = - \frac{\partial}{\partial{z_i}}( D_{1,i}^{00} \varrho(z,t)) + \frac{1}{2} \frac{\partial}{\partial{z_i}\partial{z_j}} \left( D^{00}_{2, i j} \cqstate(z,t) \right) \\
& -i[H(z), \cqstate(z,t)] \\
&+ D_0^{\alpha \beta} L_{\alpha} \cqstate(z,t) L_{\beta}^{\dag} - \frac{1}{2} D_0^{\alpha \beta} \{ L_{\beta}^{\dag} L_{\alpha}, \cqstate(z,t) \}_+  \\
&- \frac{\partial }{\partial z_{i}} \left( D^{0\alpha}_{1, i} \cqstate(z,t) L_{\alpha}^{\dag} \right)  - \frac{\partial }{\partial z_{i}} \left( D^{\alpha 0 }_{1, i} L_{\alpha} \cqstate(z,t) \right)   ,
\end{split}
\end{equation}
where 
\begin{align}
    D_0 \succeq D_1^{\dag} D_2^{-1} D_1,\,\,\,\, (\mathbb{I}- D_2 D_2^{-1})D_1 =0, \label{eq:tradeoff}
\end{align} are sufficient and necessary conditions for the dynamics to be completely positive. In the above and what follows, the $i,j,\ldots$ indices run from $1$ to $n$, while the $\alpha,\beta,\ldots$ indices run from $1$ to $p$, and we assume summation over repeated indices of either kind. Here $D_0^{\alpha \beta}$ are the elements of $D_0$, $D_{1,i}^{0 \alpha}$ are the elements of $D_1$, while $D_{1,i}^{\alpha 0}={D_{1,i}^{0 \alpha}}^*$. Additionally, $D_{1,i}^{00}$ are the elements of $D_1^C$ and $D_{2,ij}^{00}$ are the elements of $D_2$, which has the generalised inverse $D_2^{-1}$. In the positivity conditions, $A\succeq B$ is shorthand for stating that the matrix $A-B$ is positive semi-definite, and $\mathbb{I}$ is the identity operator in the space of $n\times n$ matrices. In analogy with the Lindblad equation for open quantum systems \cite{lindbald, gorini_koss_sud}, we refer to the operators $L_{\alpha}$ appearing in the master equation as \textit{Lindblad operators}; when these operators are not chosen traceless and orthogonal, the above conditions on the dynamics can be shown to be sufficient for complete positivity \cite{CQPawula}. Note that we shall also frequently refer to the first positivity condition in Equation \eqref{eq:tradeoff} as the decoherence-diffusion trade-off \cite{decodiff2}. 

The first line of \eqref{eq: continuousME} describes pure classical evolution, and more specifically a Fokker-Plank equation, which is a classical diffusion process. The second and third lines describe pure quantum evolution, which includes a pure Lindbladian term representing non-unitary evolution. Finally, the fourth line describes non-trivial CQ dynamics, where changes in the quantum state are associated to changes in the classical state. 
An early example of \eqref{eq: continuousME} where the backreaction was generated by a Hamiltonian was introduced by Diosi in \cite{diosi1995quantum}, following the bracket introduced by Aleksandrov and Gerasimenko \cite{aleksandrov1981statistical,gerasimenko1982dynamical,boucher1988semiclassical}. 

To facilitate the unravelling of Equation \eqref{eq: continuousME} into trajectories, let
 $d W_{i}$ be the standard multivariate Wiener process satisfying the standard Ito rules $ dW_i d W_j = \delta_{ij} dt$, $\ d W_i dt = 0$,  $\sigma$ be any real matrix satisfying $D_{2} =  \sigma \sigma^T$ and $\sigma^{-1}$ its generalised inverse. Additionally, define $\langle A \rangle=\tr\{A\rho_t\}$ for any operator $A$. Our main technical result is that the equations 
\begin{equation}\label{eq: unravelingCQClass}
     \begin{split}
   dZ_{t,i} =& D_{1,i}^{00} (Z_t) dt + \langle D^{\alpha 0 }_{1,i} (Z_t) L_{\alpha} + D^{0 \alpha }_{1,i}(Z_t) L_{\alpha}^{\dag} \rangle dt \\
   & + \sigma_{ij}(Z_t) d W_{j} 
 \end{split}
 \end{equation}
 and 
 \begin{equation}\label{eq: unravelingCQCQuantum}
 \begin{split}
  d \rho_t &= -i[H(Z_t), \rho_t]dt \\
  & + D_0^{\alpha \beta}(Z_t)(L_{\alpha} \rho L_{\beta}^{\dag} dt - \frac{1}{2} \{L_{\beta}^{\dag} L_{\alpha}, \rho_t \}_+ ) dt\\
   & + D_{1,j}^{\alpha 0}(Z_t) \sigma^{-1}_{ij} (Z_t) (L_{\alpha} - \langle L_{\alpha} \rangle) \rho_t d W_i \\
 & +  D_{1,j}^{ 0 \alpha}(Z_t) \sigma^{-1}_{ij}(Z_t)  \rho_t (L_{\alpha}^{\dag} - \langle L_{\alpha}^{\dag} \rangle)  d W_i
 \end{split}
 \end{equation}
define an unravelling of the master Equation \eqref{eq: continuousME} for which $\rho(t|\sigma\{Z_s\}_{s\leq t})=\rho_t$. In other words, once averaged over the noise process, the equations \eqref{eq: unravelingCQClass} and \eqref{eq: unravelingCQCQuantum} define the most general allowed master equation for the combined classical-quantum state $\cqstate(z,t)$; and it is dynamics such that an observer with access to the full classical trajectory up to time $t$ may deduce $\rho_t$ from the changes $dZ_t$ in the classical degrees of freedom. 

To prove that the unravelled dynamics is equivalent to the master equation for the classical-quantum state, we follow the same steps as in the derivation of equation \eqref{eq: nonlinearME}. The key difference is that when using the chain rule, one must expand to second order in terms containing $dW_i$, and use the Ito rules. Doing so, one finds that the appearance of additional terms are sufficient to cancel the terms in the standard semi-classical equations that previously prevented their linearity, with the final master equation being given by the linear equation \eqref{eq: continuousME}. To prove that the dynamics satisfies $\rho(t|\sigma\{Z_s\}_{s\leq t})=\rho_t$, one may explicitly rewrite \eqref{eq: unravelingCQCQuantum} such that the noise terms $dW_i$ are replaced with terms containing $dZ_i$. Since the steps involved are rather long, we leave the full proofs, along with an alternate vectorised notation for the dynamics, to Appendix \ref{sec: unravelingCont}. 

Although the expectation values $\langle L_\alpha^{(\dag)}
\rangle $ appear in the equations for both $Z_t$ and $\rho_t$, the reasons they appear in each are distinct. In the quantum dynamics of Equation \eqref{eq: unravelingCQCQuantum}, the appearance of the expectation values causes the resulting map to be non-linear in $\rho_t$. Although non-linear maps are problematic when interpreted as evolution maps alone \cite{schmid2019initial}, in this case the map is describing both the evolution and the result of conditioning on the classical trajectory, as indicated by the relation $\rho_t=\rho(t|\sigma\{Z_s\}_{s\leq t})$. In Appendix \ref{sec: joint} we show explicitly that the equivalent dynamics of a joint classical-quantum state is non-linear solely due to normalisation, and thus the expectation values appearing in Equation \eqref{eq: unravelingCQCQuantum} appear due to conditioning, just as in the unravellings of the GKSL equation \cite{breuer2002theory}. By contrast, the appearance of expectation values in \eqref{eq: unravelingCQClass} expresses the uniquely quantum feature of the dynamics that the drift of the classical system is generically unknown, even if the quantum system is in a pure state.

One would be forgiven in thinking that the dynamics in Equation \eqref{eq: continuousME}, and thus the equivalent dynamics of Equations \eqref{eq: unravelingCQClass} and \eqref{eq: unravelingCQCQuantum}, lead to a loss of quantum information due to the presence of the decoherence terms with coefficients $D_0$. However for an initially pure quantum state, we find that when the decoherence-diffusion trade-off is saturated (i.e.  $D_0 =  D_1^\dag (\sigma \sigma^T)^{-1} D_1 $), the quantum state remains pure i.e.  $d\tr\{\rho_t ^2\}= 0$. That is, there is no loss of quantum information. We can see this more explicitly via the unique pure state unravelling
\begin{equation}\label{eq: unravelingCQPsi}
\begin{split}
     & d |\psi \rangle_t = -i H(Z_t)|\psi \rangle_t dt  \\
     & + D_{1,j}^{\alpha 0}(Z_t) \sigma^{-1}_{ij} (Z_t) (L_{\alpha} - \langle L_{\alpha} \rangle) |\psi \rangle_t d W_i\\
     & - \frac{1}{2} D_0^{\alpha \beta}(Z_t) (L_{\beta}^{\dag} - \langle L_{\beta}^{\dag} \rangle) (L_{\alpha} - \langle L_{\alpha} \rangle)|\psi\rangle_t dt \\
     & + \frac{1}{2} D_0^{\alpha \beta}(Z_t)( \langle L^\dag_\beta\rangle L_\alpha -  \langle L_\alpha \rangle  L^\dag_\beta) | \psi \rangle_t dt
\end{split}
\end{equation}
which using the standard Ito rules 
\begin{equation}\label{eq: stateVectorEvolution}
    d \rho_t = d|\psi \rangle_t \langle \psi |_t + |\psi \rangle_t d \langle \psi |_t + d |\psi \rangle_t d\langle \psi |_t
\end{equation} 
is equivalent to \eqref{eq: unravelingCQCQuantum} when the decoherence diffusion trade-off is saturated. Since $\rho_t=\rho(t|\sigma\{Z_s\}_{s\leq t})$, this implies that there is a unique quantum state conditioned on the classical trajectory that remains pure
at all times. Thus, despite loss of predictability in the classical degrees of freedom, our theory, and thus general hybrid classical-quantum theories saturating the trade-off, do not exhibit any loss of predictability in the quantum degrees of freedom when conditioned on the classical trajectory.

When the trade-off is not saturated, the quantum state conditioned on the classical trajectory will generically demonstrate decoherence e.g. due to an external quantum or classical environment. Inspired by the purely quantum case, where mixed quantum states can be purified by considering states in a larger quantum environment, we prove in Appendix \ref{sec: temple} by explicit construction that one may always purify the dynamics \eqref{eq: unravelingCQClass} and \eqref{eq: unravelingCQCQuantum} by an equivalent dynamics in an enlarged classical phase space or quantum Hilbert space such that the decoherence-diffusion trade-off is saturated. One can therefore always consider classical-quantum dynamics in terms of pure quantum states and points in phase space, in much the same way that quantum dynamics may always be considered to be pure in some larger Hilbert space.

\section{Healed semi-classical dynamics} \label{sec: healed}
The general dynamics for the classical degrees of freedom $Z_t$ and the quantum state $\rho_t$, given by \eqref{eq: unravelingCQClass} and \eqref{eq: unravelingCQCQuantum}, provide a general class of dynamics to describe semi-classical systems. Using the freedom in the $D$ matrices and $H$, one may attempt to construct sensible semi-classical dynamics phenomenologically by fitting predictions of these models to data. An alternative approach is to find a dynamics that explicitly resembles the standard semi-classical dynamics of \eqref{eq: hamSemiClass} and \eqref{eq: hamSemiQuant}, which we will now turn to in this section. 

Starting with Equations \eqref{eq: unravelingCQClass} and \eqref{eq: unravelingCQPsi}, we take the pure classical part of the drift to be generated by a classical Hamiltonian $H_C(z)$. For the interaction terms, one can use the freedom in the choice of Lindblad operators to pick $L_\alpha = \{Z_\alpha,H_I\}$, where $H_I(z)$ is an interaction Hamiltonian, and then set $D_{1,i}^{0 \alpha} = \frac{1}{2} \delta_i^{\alpha}$ – this corresponds to choosing backreaction given by the Alexandrov bracket \cite{aleksandrov1981statistical,gerasimenko1982dynamical,boucher1988semiclassical,diosi1995quantum}. This fixes the decoherence term to be ${D_0=\frac{1}{4}(\sigma \sigma^T)^{-1}}$, since by assumption we assume the decoherence-diffusion trade-off is saturated. Finally, we ensure that that the first term of \eqref{eq: unravelingCQPsi} has a Hamiltonian picked that coincides with the classical-quantum Hamiltonian $H(z)=H_C(z)\mathds{1}+ H_Q + H_I(z)$. Using this definition of $H(z)$ to simplify the various terms, we arrive at a set of equations that we dub ``the healed semi-classical equations'', given as 
\begin{align} \label{eq: healedSemiclass}
    &dZ_{t,i} =  \langle \{Z_{i}, H \} \rangle dt + \sigma_{ij} d W_j \\
    & \nonumber\\
\label{eq: healedSemiQuant}
     d |\psi \rangle_t &= -i H|\psi \rangle_t dt   + \frac{1}{2} \sigma^{-1}_{ij} \{Z_j, H-\langle H \rangle \}  |\psi \rangle_t d W_i \nonumber\\
      & -\frac{1}{8} \sigma_{ij}^{-1} \sigma_{ik}^{-1} \{Z_j, H-\langle H\rangle \} \{Z_k, H-\langle H \rangle \} |\psi \rangle_t dt 
\end{align}
where in the above $\sigma$ may be any real matrix such that 
\begin{equation}
    (\mathbb{I}-\sigma \sigma^{-1})\{z, H\}=a(z) \mathds{1},
\end{equation} for a real vector $a(z)\in \mathbb{R}^n$; this is a sufficient and necessary condition for complete positivity. As before, while $\sigma$ may have arbitrary dependence on $z$, it cannot have dependence on the quantum state itself. For a given initial quantum state $|\psi_i\rangle$ and classical state $z_i$, these coupled stochastic differential equations determine the probability of ending up in any final pair of states $z_f$ and $|\psi_f\rangle$.  An early example of this dynamics for the special case of linear, constant force coupling between two particles, one classical and one quantum, was described in \cite{DiosiHalliwel}. Here we present the general form, and our healed version of the semi-classical Einstein equation may be found in Appendix \ref{sec: fullCQ}. 

Written in the above form, the differences with the standard semi-classical equations are clear. The classical evolution of the healed semi-classical Equation \eqref{eq: healedSemiclass} takes the form of the standard semi-classical Equation \eqref{eq: hamSemiClass}, i.e. backreaction given by an expectation value, but with an additional diffusive noise term. Similarly, the quantum evolution of the healed semi-classical Equation \eqref{eq: healedSemiQuant} takes the form of the standard semi-classical Equation \eqref{eq: hamSemiQuant} i.e. pure unitary evolution, but with an additional stochastic term that tends to drive the quantum state towards a joint eigenstate of the operators $\{z_i, H\}$ and $H$, where one exists, and an additional deterministic term than ensures that $|\psi\rangle_t$ remains normalised at all times \cite{GisinMeas}. Despite the appearance of an expectation value in the backreaction drift term, the joint dynamics of these coupled equations gives statistics for $Z_t$ as if the classical system were diffusing around a force given by a random eigenstate of the operators $\{z_i, H\}$, just as the left hand side of Figure \ref{fig:notthismain} depicts. The free parameters of the model $\sigma_{ij}$ determine both the rate at which the quantum state evolves to an eigenstate and the rate of diffusion of the classical system – that these two rates are explicitly inversely related is the expression of the decoherence-diffusion trade-off. The other positivity condition of \eqref{eq:tradeoff} appears as the condition ${(\mathbb{I}-\sigma \sigma^{-1})\{z, H\}=a(z) \mathds{1}}$ for some $a(z)$, which ensures that no combination of the classical degrees of freedom can be constructed such that a classical variable has a drift depending on a quantum expectation value without an associated noise term. When the back-reaction is zero, i.e. $H_I(z)=0$, the positivity condition is satsified for all $\sigma$, and thus $\sigma$ may be taken to zero. In this limit, the equations reduce to uncoupled deterministic Hamiltonian classical and quantum mechanics.

\begin{figure}[]
    \centering
    \includegraphics[width=8.8cm,trim={0.3cm 2.5cm 2.5cm 2.2cm},clip]{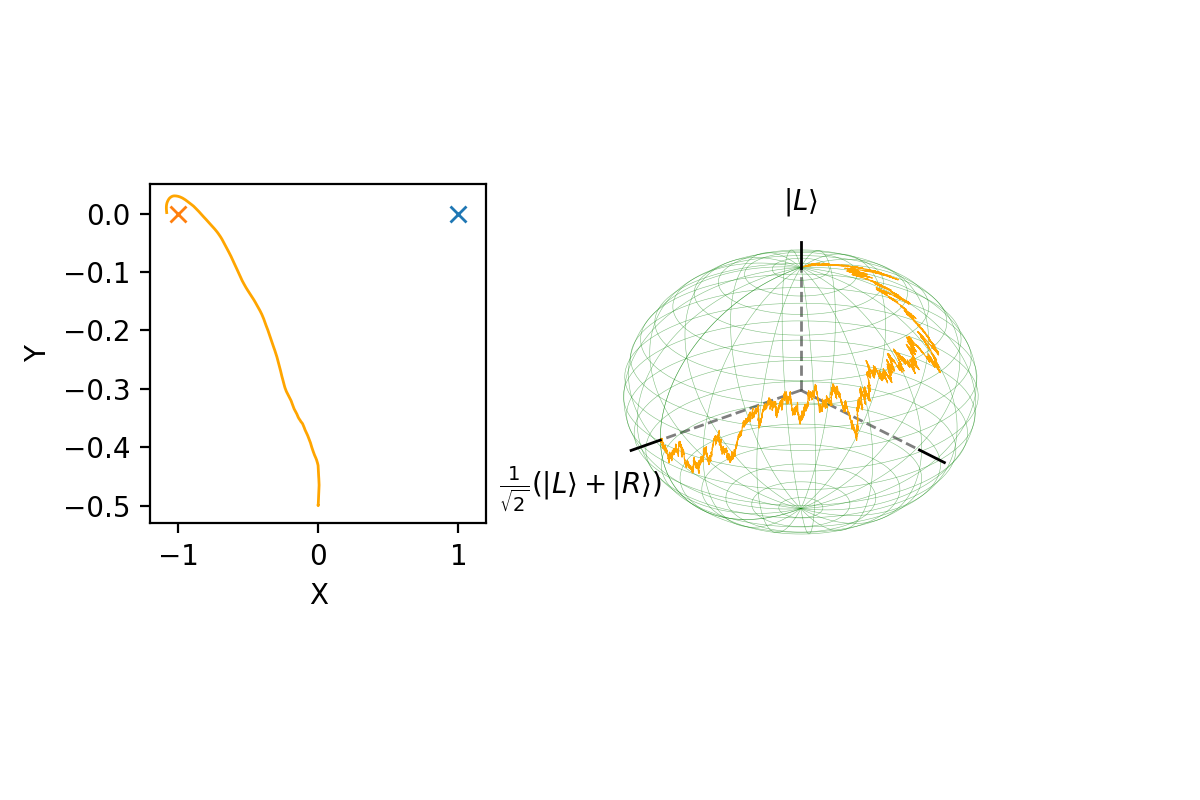}
    \caption{\footnotesize{A numerically simulated classical-quantum trajectory for a toy model of a classical test particle moving in the Newtonian potential sourced by a mass in a spatial superposition of left $|L\rangle$ and right $|R\rangle$ states (positions marked in space by crosses). Initially starting at rest at $X=0$ and $Y=-0.5$, the test particle is seen to eventually fall toward the mass being on the left, with the small initial motion toward the centre of the two possible locations accounted for by the diffusion in momentum. This should be contrasted with the right hand panel of Figure \ref{fig:notthismain} depicting the dynamics using the standard semi-classical equations. See Appendix \ref{sec: toy} for for more information.}}
    
    \label{fig:single_planet_traj}
\end{figure}

To make the above discussion more concrete, consider the numerically simulated dynamics shown in Figures \ref{fig:single_qubit_traj} and \ref{fig:single_planet_traj}, of a qubit interacting with a 1D particle, and a toy model of a test particle moving in the Newtonian potential of a mass in superposition. Here the trajectories in phase space are shown on the left, while on the right the trajectory in Hilbert space is represented by a path on the surface on the Bloch sphere. For both, we see that the classical trajectories correspond to the motion expected if the force on the classical system were determined by an eigenvalue of $\{p_i,H\}$, but with additional diffusion around this. The quantum trajectories are correlated with these classical trajectories, such that when the particle's momentum has increased significantly, or when the test mass has moved significantly towards the mass on the left, the corresponding quantum state is also in the corresponding eigenstate with high probability. A given change in the classical system is only significant if it is large compared to the background classical noise $\sigma$; this indicates why the changes in the quantum degree of freedom are inversely proportional to the noise strength in the classical system.

The full details of these models appear in Appendix $\ref{sec: toy}$, where we simulate a number of simple toy models that arise as special cases of this general Hamiltonian dynamics, including a toy model for vacuum fluctuations sourcing the expansion rate in the early universe.

\section{Regime of validity of the standard {semi-classical} equations} 

Having derived a general form of consistent semi-classical dynamics, we may test the regime of validity of the standard semi-classical equations \eqref{eq: hamSemiClass} and \eqref{eq: hamSemiQuant} by studying when they can be approximately recovered from the consistent dynamics of \eqref{eq: healedSemiclass} and \eqref{eq: healedSemiQuant}.

The first important observation to make is that that the standard semi-classical equations cannot be derived as a limit of the healed ones. Specifically, while the free parameters contained in the matrix $\sigma$ of the healed semi-classical equations may be varied such that either \eqref{eq: healedSemiclass} or \eqref{eq: healedSemiQuant} approximately takes the form of the respective standard semi-classical equation, the appearance of both $\sigma$ and $\sigma^{-1}$ prevent the recovery of both equations in any limit, with either the diffusion or decoherence necessarily becoming large as $\sigma\rightarrow \infty$ or $\sigma\rightarrow 0$. In the special case that $\sigma=0$, the positivity condition reduces to ${\{z, H\}}=a(z)\mathds{1}$ for some $a(z)\in \mathbb{R}^n$ i.e. that the classical-quantum Hamiltonian is of the form $H(z)=H_C(z)\mathds{1} + H_Q$, meaning that the quantum backreaction on the classical system must also be zero. As should be expected, the failure to recover the standard semi-classical equations from their healthier versions unless the quantum backreaction is zero is consistent with the analysis in Sec. \ref{sec: standard},  where the dynamics induced on the classical-quantum state was found only to be linear provided $H_I(z)$ was zero.

The standard semi-classical equations are thus inconsistent if applied to all states. However, using Equations \eqref{eq: healedSemiclass} and \eqref{eq: healedSemiQuant}, we can find a regime in which the standard semi-classical equations are approximately valid for a given initial quantum state $|\phi\rangle$. Specifically, we will ask which initial states $|\phi\rangle$ and timescales $\tau$ the healed semi-classical equations can be approximated by the standard ones i.e. when the additional terms corresponding to diffusion and decoherence have a negligible impact on observations. 

For the classical degrees of freedom to appear to evolve deterministically, one should consider timescales much larger than the noise to signal ratio of the system, such that any noise fluctuations are removed by coarse-graining in this time window \cite{primak2005stochastic}. Provided the classical system evolves slowly, and is only weakly affected by the quantum one, we may approximate the signal provided by the $i$th degree of freedom of the classical system as the square of the initial drift $\langle \phi| \{Z_i,H\}|\phi \rangle$, and the noise in this degree of freedom as $(\sigma \sigma^T)_{ii}$, the initial value of this component of the diffusion matrix. For the dynamics of \eqref{eq: healedSemiclass} to reduce to \eqref{eq: hamSemiClass}, one must consider timescales $\tau$ large compared to the largest possible ratio of the diffusion and the square of the drifts of all the classical phase space degrees of freedom $i$ i.e.
\begin{equation} \label{ineq: timescale_lower}
    \tau \gg \max_{i} \frac{(\sigma \sigma^T)_{ii}}{\langle \phi |\{Z_i,H\}| \phi \rangle^2}.
\end{equation} On the other hand, for the quantum system to appear to evolve unitarily, one must consider timescales very short compared to the decoherence time of the quantum system. The decoherence rate may be computed for an initially pure quantum state via the sum of the variances of the diagonal Lindblad operators $\bar{L}_i$ acting on the system  \cite{beau2017nonexponential}, which in our case take the form $\bar{L}_i=\sigma^{-1}_{ij}\{Z_j,H\}$. This means that \eqref{eq: healedSemiQuant} only reduces to \eqref{eq: hamSemiQuant} on timescales
\begin{equation} \label{ineq: timescale_upper}
    \tau \ll \frac{1}{\sum_{ijk} \sigma^{-1}_{ij}\textrm{cov}_{|\phi\rangle} (\{z_j,H\},\{z_k,H\})\sigma^{-T}_{ki}}
\end{equation} where here $\textrm{cov}_{|\phi\rangle}(A,B)=\langle AB \rangle-\langle  A  \rangle \langle B \rangle $ with expectation values taken with respect to $|\phi\rangle$. We thus see that simultaneously requiring the decoherence and diffusion to be negligible both upper and lower bounds the timescale $\tau$ for which the standard semi-classical equations approximate the consistent dynamics of the healed semi-classical equations, for a given initial quantum state $|\phi\rangle$.

The above conditions on initial quantum states $|\phi \rangle$ and timescales $\tau$ can be adapted to prove a concise necessary condition on the regime of validity of the standard semi-classical equations. Firstly, we note that \eqref{ineq: timescale_lower} implies that $\tau$ must be much larger than each of the $i$ noise to signal ratios, and therefore that $\langle \phi |\{Z_i,H\}| \phi \rangle^2$  must be much greater than the ratio of $(\sigma \sigma^T)_{ii}$ to $\tau$. Secondly, we note that the denominator of \eqref{ineq: timescale_upper} may be rewritten as a trace of a positive semi-definite matrix $A=\sigma^{-1}C_{|\phi \rangle}\sigma^{-T}$, where $C_{|\phi \rangle}$ is the matrix of the covariances introduced previously. Using the inequality $\tr A \geq (v^T A v)/(v^T v) $, which holds for an arbitrary vector $v$, it follows from \eqref{ineq: timescale_upper} that $\tau$ must be much smaller than $(v^T v)/(v^T A v)$ for all $v$. Taking $v$ to be given by each of the columns of $\sigma$, and using the positivity condition ${(\mathbb{I}-\sigma \sigma^{-1})\{z, H\}}=a(z) \mathds{1}$ to replace $\sigma \sigma^{-1} C_{|\phi \rangle} \sigma^{-T} \sigma^{T}$ with $C_{|\phi \rangle}$, one finds that $\tau$ is much smaller that each of the ratios of $(\sigma \sigma^T)_{ii}$ to the $i$th diagonal of the covariance matrix $C_{|\phi\rangle}$ i.e. the variance, which we denote $\textrm{var}_{|\phi\rangle}(\{z_i,H\})=\langle \phi| \{z_i,H\}^2| \phi \rangle-\langle \phi| \{z_i,H\}| \phi \rangle^2$. Rearranging this final inequality and combining with the ones derived from \eqref{ineq: timescale_lower}, we arrive at the following set of inequalities 
\begin{equation} \label{ineq: timescale_sando}
    \langle \phi |\{z_i,H\} | \phi \rangle^2 \gg \frac{(\sigma \sigma^T)_{ii}}{\tau} \gg \textrm{var}_{|\phi\rangle}(\{z_i,H\}),
\end{equation} which must hold for each $i$ in order for the standard semi-classical equations \eqref{eq: hamSemiClass} and \eqref{eq: hamSemiQuant} to be a valid approximation to \eqref{eq: healedSemiclass} and \eqref{eq: healedSemiQuant}. We thus see that for there to exist any timescale $\tau$ for which the standard semi-classical equations are a good approximation to a consistent semi-classical dynamics, the force on the classical system, either from quantum backreaction or internal classical dynamics, must be much greater than its variance with respect to the quantum state.  While this condition has been previously postulated \cite{Kuo:1993if}, the analysis above provides a rigorous derivation, by requiring the standard semi-classical dynamics to be a valid approximation of a consistent semi-classical theory. Moreover, for a given quantum state $|\phi\rangle$ that satisfies $\langle \phi |\{z_i,H\} | \phi \rangle^2 \gg  \textrm{var}_{|\phi\rangle}(\{z_i,H\})$, the full set of inequalities \eqref{ineq: timescale_sando} gives the timescale over which the dynamics will be valid, for a given diffusion/decoherence rate controlled by $\sigma$.

In the limiting case that the state $|\phi\rangle$ is an eigenstate of the operators $\{z_i,H\}$, the variance vanishes, and thus the standard semi-classical dynamics are valid approximation for all times provided the diffusion $\sigma \sigma^T$ is sufficiently small. In the Newtonian limit of gravity, the interaction is dominated by the mass density $\frac{\partial H}{\partial \Phi} =\hat{m}(x)$ and we see that the standard semi-classical equations are exactly valid only when the quantum state is in an approximate eigenstate of the mass density operator, which excludes macroscopic superpositions, as well as states which are spatially entangled: essentially the quantum state of matter must be approximately classical \cite{Ford:1982wu, Kuo:1993if, Hu:2000ns}. 

However, while the above analysis implies that there can exist certain states and timescales over which the standard semi-classical approach is a reasonable approximation to a consistent semi-classical dynamics, it must be emphasised that in practice the regime of validity may be very limited. The healed equations of \eqref{eq: healedSemiclass} and \eqref{eq: healedSemiQuant} thus provide a semi-classical dynamics that extends the regime of validity of the standard semi-classical dynamics to arbitrary states and timescales.

An important example of where the healed semi-classical dynamics extends the regime of validity of the standard semi-classical equations, without considerable increased technical difficulty, is in the low noise, $\sigma\rightarrow0$, limit of the theory. Here, one considers timescales long compared to the diffusion in the theory, such that the inequality of \eqref{ineq: timescale_lower} holds, and the classical equations can be approximated by \eqref{eq: hamSemiClass}. However, rather than restricting to specific states and timescales such that the decoherence is negligible, we instead keep the additional $\sigma^{-1}$ dependent terms in \eqref{eq: healedSemiQuant} that do not appear in \eqref{eq: hamSemiQuant}. In the limit that $\sigma \rightarrow 0$, the quantum dynamics causes the quantum state to almost instantaneously evolve, with probabilities given by the Born rule, to an eigenstate of the operators $\{z_i,H\}$ \cite{GisinMeas,2006Jacob}. The classical evolution is thus well approximated by conditioning on eigenstates of the quantum state decohered in this basis, and then evolving according to classical equations of motion. This is in fact the way in which the semi-classical Einstein equations are often used in practice to deal with classical mixtures (see also Appendix \ref{sec:superluminal}) – here we see that this use of them is a limiting case of the healed semi-classical equations when the diffusive noise in the classical system is negligible. In this limit, the resulting classical system is still described by a probability distributions over final states, but this distribution is entirely due to the probability distribution over eigenstates of $\{z_i,H\}$ provided by the decohered quantum state.  

While the low noise limit of the healed semi-classical dynamics may be a valid regime in which to study semi-classical physics, more generally one would like to understand what happens while the quantum state still has coherence. Here, the final probability distribution of the classical system is due to both the initial quantum state and diffusion in the classical system itself. In this case $\sigma$ is finite, and the full machinery presented thus far must be used.

\section{Comparison with measurement and feedback} 
We can also compare our result to previous methods of generating consistent classical-quantum dynamics using continuous measurement and feedback approaches \cite{DiosiHalliwel, 2016Tilloy,kafri2014classical}. In these approaches, the classical degree of freedom is sourced by the outcomes of a continuous measurement, and by construction such approaches are completely positive and lead to consistent coupling between classical and quantum degrees of freedom. The stochasticity of the dynamics due to the continuous measurement, and the non-linearity due to the state update rule, mean the dynamics of \cite{DiosiHalliwel, 2016Tilloy,kafri2014classical} take a similar form to Equations \eqref{eq: healedSemiclass} and \eqref{eq: healedSemiQuant}. However, it is worth noting some differences between the various approaches based on continuous measurement, and the one we have presented so far. Firstly, the dynamics we present allow for the classical degrees of freedom to be independent of the quantum degrees of freedom, and have their own dynamics, described via the purely classical evolution term $D_{1,i}^{00}$. This allows us to apply the dynamics to more complex CQ scenarios where there is self-gravitation, as well as when the interaction is generated by a non-linear Hamiltonian $H(z)$. As a result, the dynamics presented here, while autonomous on both the classical and quantum system, can be non-Markovian on the classical and quantum systems alone. It therefore does not always reduce to pure Lindbladian evolution on the quantum system, such as in \cite{2016Tilloy, tilloy2017principle, kafri2014classical}.
 
 Secondly, we have taken the dynamics on the phase space degrees of freedom to be continuous. In the measurement and feedback approaches of \cite{2016Tilloy,kafri2014classical}, the classical degrees of freedom evolve discontinuously because the classical coordinate is directly sourced by the outcome $J_i$ of a continuous measurement which is a discontinuous stochastic random variable. To obtain continuous classical degrees of freedom, one can instead source the conjugate momenta of the canonical coordinates via the measurement signal $J_i dt $. This is the approach taken in \cite{DiosiHalliwel}, which leads to a special case of our dynamics. In Appendix \ref{sec: contMeasurement}, we show that our dynamics of Equations \eqref{eq: unravelingCQClass} and \eqref{eq: unravelingCQCQuantum} have an equivalent description in terms of a generalisation of the procedure given in \cite{DiosiHalliwel}, where an auxiliary classical degree of freedom is sourced by the measurement signal of a continuous measurement, and we also allow for auxiliary variable to have its own purely classical dynamics. In this sense, Equations \eqref{eq: unravelingCQClass} and \eqref{eq: unravelingCQCQuantum} form a complete parameterization for continuous measurement based classical-quantum control where one also allows for continuous control on the classical system, and are similar to measurement based feedback equations familiar in quantum control \cite{2006Jacob, Wiseman_2001}. It would be interesting to find a complete parameterization of the dynamics in the discontinuous case.

Finally, we note that the equivalence of classical-quantum dynamics with continuous measurement and feedback also has practical significance for numerical studies of classical-quantum systems. Given the long history of numerical studies of diffusive unravellings and quantum continuous measurement theory \cite{gisin1992quantum,Wiseman_2001,breuer2002theory,wiseman_milburn_2009}, these earlier works may be used to benchmark numerical methods for studying classical-quantum systems. Indeed, as a basic check of the simulations presented in Section \ref{sec: healed}, it is straightforward to compare qualitatively the results to numerical studies of diffusive unravellings for qubit models to see similar rotation around the Bloch sphere and stochastic collapse to the $z$-axis \cite{gisin1992quantum,Wiseman_2001}, which is reasonable given the similarity of the underlying dynamics and discretisation  schemes involved (see Appendix \ref{sec: toy} for further details). Moving forward, understanding which existing methods and models in continuous measurement theory may be used and adapted to study classical-quantum dynamics, allowing numerical simulations to be improved and properly benchmarked, is an important open question from our current work.

\section{Classical-quantum gravity} \label{sec: CQgravity}
Thus far we have only considered continuous classical degrees of freedom. In Appendix \ref{sec: field} we discuss how one can formally arrive at dynamics for fields -- the result is the same but to replace quantities with their local counterparts and derivatives with functional derivatives. Effectively, the spatial coordinate $x$ acts like an index of the Lindblad operators and the matrices $D_n$.

Though our goal here is not to reproduce a fully covariant semi-classical description of quantum gravity, but rather, provide a framework to clarify some of the major issues of that approach, we conclude with a brief discussion of the full gravitational context. 
Models of classical-quantum Newtonian gravity were introduced in \cite{2016Tilloy,kafri2014classical} using a measurement-feedback approach. Classical-quantum dynamics in the full gravitational setting has been studied in \cite{oppenheim_post-quantum_2018, Oppenheim:2020ogy,covariantPI}. The idea, introduced in \cite{oppenheim_post-quantum_2018}, was to take the classical degrees of freedom to be given by the Riemmanian $3$ metrics (on some $3$ surface $\Sigma$) and their conjugate momenta $z= (g_{ij}, \pi^{ij})$. One then considers completely positive dynamics, depending on some lapse $N$ and shift $N^i$, which maps hybrid states $\cqstate(g, \pi,t)$ onto themselves, describing a geometrodynamic picture of classical-gravity interacting with quantum matter. One can also consider the lapse and shift and their conjugate momenta to be part of the phase space, in which case they enter into the Poisson bracket. While this changes nothing in the purely classical case, it offers some advantages in the CQ case. Here, by (formally) studying the unraveling of the dynamics, for each realization of the noise process, we now have entire trajectories for each of the variables $(g_{ij}, \pi^{ij}, N, N^i)$ each associated to a quantum state, $\rho(t | g_{ij}, \pi^{ij}, N, N^i)$. This allows us to define a tuple $(g_{\mu \nu}, \rho_{\Sigma_t}(t))$ via the ADM embedding
\begin{equation} 
\begin{split}
\label{eq: admdecom}
& g_{\mu \nu} dx^{\mu} dx^{\nu}  = -N^2(t,x)dt^2 \\
&+ g_{ij}(t,x)(N^i(t,x) dt + dx^i)(N^j(t,x)dt + dx^j). 
\end{split}
\end{equation}
This associates to each trajectory a 4-metric and quantum state on a 1-parameter family of hypersurfaces $\Sigma$. The unraveling thus provides a method to study the dynamics of classical gravity interacting with quantum matter. 

Taking the pure dynamics to be local and Hamiltonian, in the sense of Equations \eqref{eq: healedSemiclass}, \eqref{eq: healedSemiQuant}, in Appendix \ref{sec: field} we find the dynamics
\begin{equation} \label{eq: healedSemiclassField}
    G_{ij} = 8 \pi G \langle T_{ij}[g, \pi]\rangle + \sigma_{ij}^{kl}[g,\pi] \dxi_{kl},
\end{equation}
where $\dxi_{kl}$ is a white noise process. The evolution of the quantum state is given by Equation \eqref{eq: healedSemiQuant} where $H_I(z)$ is the matter Hamiltonian. The construction of a classical-quantum theory with the same number of degrees of freedom as GR amounts to constructing the hybrid versions of the gravitational constraints which are the $G_{00}$ and $G_{0i}$ components of the Einstein tensor. Hybrid classical-quantum constraints have been studied in the context of master Equations \cite{oppenheim_post-quantum_2018, Oppenheim:2020ogy} but are currently not well understood. In \cite{covariantPI, pathIntegralLong} we introduce a diffeomorphism covariant and invariant theory of classical-quantum gravity from a path integral perspective, based on the trace of Einsten's equation. While it does not give full general relativity, it serves as a proof of principle that classical-quantum theories may be made diffeomorphism invariant and may lead to insight into the constraints. In Appendix \ref{sec: field} we include a discussion of interacting classical-quantum gravity. Importantly, the decoherence-diffusion trade-off can be used to experimentally test for theories with a fundamentally classical gravitational field since they necessarily lead to diffusion in the gravitational potential and decoherence of masses in spatial superposition \cite{decodiff2}.

On a related note, the form of the classical evolution equation \eqref{eq: healedSemiclassField} looks similar to a Markovian version of the non-Markovian Einstein-Langevin equation \cite{CalzettaHu, HuEL,Ros97, Martin:1998nc,Hu:2008rga}. This is the central object of study in ``stochastic gravity" \cite{CalzettaHu, HuEL,Ros97, Martin:1998nc,Hu:2008rga} aimed at incorporating higher order corrections to Einstein's equations sourced by the quantum stress energy tensor. Such corrections were originally motivated by studying the interaction of two linear quantum systems via a path integral approach, integrating out one of the quantum systems and looking at the $\hbar \to 0 $ dynamics of the remaining system. The Einstein-Langevin equation is believed to be valid whenever the dynamics can be approximated by correlation functions which are second order. The dynamics we introduce provides a semi-classical regime which goes beyond this, since we have arrived at a consistent semi-classical picture which gives rise to consistent dynamics on \textit{any} quantum state: this includes quantum states in macroscopic superposition, which are not approximated well by second order correlation functions and for which the Einstein-Langevin equation fails to be a good approximation \cite{Hu:2008rga}.

\section{Discussion} The equations we find parameterise the general form of completely-positive, linear, autonomous and continuous classical-quantum dynamics in terms of trajectories in phase space and Hilbert space. Given the initial motivation was to arrive at a healthier theory of semi-classical gravity, it is worth considering carefully when we expect these assumptions to hold. If gravity were to be fundamentally classical, then these assumptions are reasonable: the assumptions of complete-positivity and linearity are necessary for sensible predictions for all initial classical and quantum states; the assumption of autonomous dynamics is reasonable for any theory viewed as fundamental; and the assumption of continuous classical trajectories is necessary for the dynamics to describe probability distributions over spacetimes. Viewed in this way, one expects that the field theoretic versions of Equations \eqref{eq: unravelingCQClass} and \eqref{eq: unravelingCQCQuantum} (Appendix \ref{sec: field}) provide a template to construct consistent CQ theories of gravity. 

The theory presented here may also be used to study regimes in which a degree of freedom, such as gravity, is effectively classical. In general, effective theories need not necessarily satisfy all of the assumptions, at least exactly or for all times. For instance, if one allows for the non-Markovian evolution that generically arises in the study of open quantum systems, we necessarily violate the assumption of autonomous dynamics. In this case, a time-local non-Markovian theory takes the same form as Equation \ref{eq: continuousME}, but without the requirement for the decoherence-diffusion trade-off to hold for all times \cite{oppenheim_post-quantum_2018} (see also Appendix \ref{sec: MarkovianApp}). Alternatively, one may construct dynamics which although are not completely positive on all initial classical distributions, are completely positive on those permissible by a quantum theory e.g. for which $\Delta q \Delta p\geq \frac{\hbar}{2}$ and so resemble Equation \ref{eq: continuousME}, but without the same positivity requirements \cite{diosi2000quantum}. However, while such theories may be useful as effective theories, they do not appear to be compatible with the prediction of observable classical trajectories, as one should expect for an effective theory of classicality. In \cite{layton2024classical} a general framework for effective classicality is developed, in which well-defined classical trajectories emerge in ``the classical-quantum limit" of a quantum dynamics. The resulting dynamics arise as a special case of Equations \eqref{eq: unravelingCQClass} and \eqref{eq: unravelingCQCQuantum}, demonstrating that the framework presented here may indeed be applied as an effective theory. 

Since the current theory can be interpreted as an effective one \cite{layton2024classical}, one may also expect the dynamics presented here to apply to situations beyond the study of gravity. For example, one may hope that our dynamics in Equations \eqref{eq: healedSemiclass} and \eqref{eq: healedSemiQuant} could be applied to the study of molecular dynamics, where heavy nuclear degrees of freedom are approximated as classical, while the electronic degrees of freedom remain quantum \cite{tully1990molecular}. However, such an approximation is only likely to be valid when the effective classical degrees of freedom have well-defined trajectories, which is only expected to arise when the decoherence from the environment on these degrees of freedom is strong \cite{layton2024classical}. One must therefore must be careful in applying the current framework as a general method of approximating quantum dynamics.

A consistent treatment of classical-quantum trajectories may shed light on some of the open problems in semi-classical physics. 
Of potential interest is understanding the role of vacuum fluctuations in cosmology and structure formation. Since we wish to investigate the role that vacuum fluctuations play in density inhomogeneity, this is a regime in which the semi-classical Einstein Equation \eqref{eq: semiclassEinstein} cannot be used. In practice,
researchers consider situations in which the density perturbations have decohered \cite{Kiefer:2006je, KieferPolarski,Kiefer:1998qe,Halliwell:1989vw, starobinsky86,starobinsky1994equilibrium, Polarski:1995jg}, so that they can condition on their value and feed this into the Friedman-Robertson-Walker equation governing the expansion of the local space-time patch \cite{Sasaki,Guth1,Guth2}. Such an approach is inconsistent with the semi-classical Einstein Equation \cite{Perez:2005gh} (see also Appendix \ref{sec:superluminal}). As already discussed, this procedure can in fact be understood as the low noise, $\sigma\rightarrow 0$, limit of the healed semi-classical equations \eqref{eq: healedSemiclass} and \eqref{eq: healedSemiQuant}. 

\begin{figure}[]
    \centering
    \includegraphics[width=8.8cm,trim={0.3cm 1cm 2cm 2.2cm},clip]{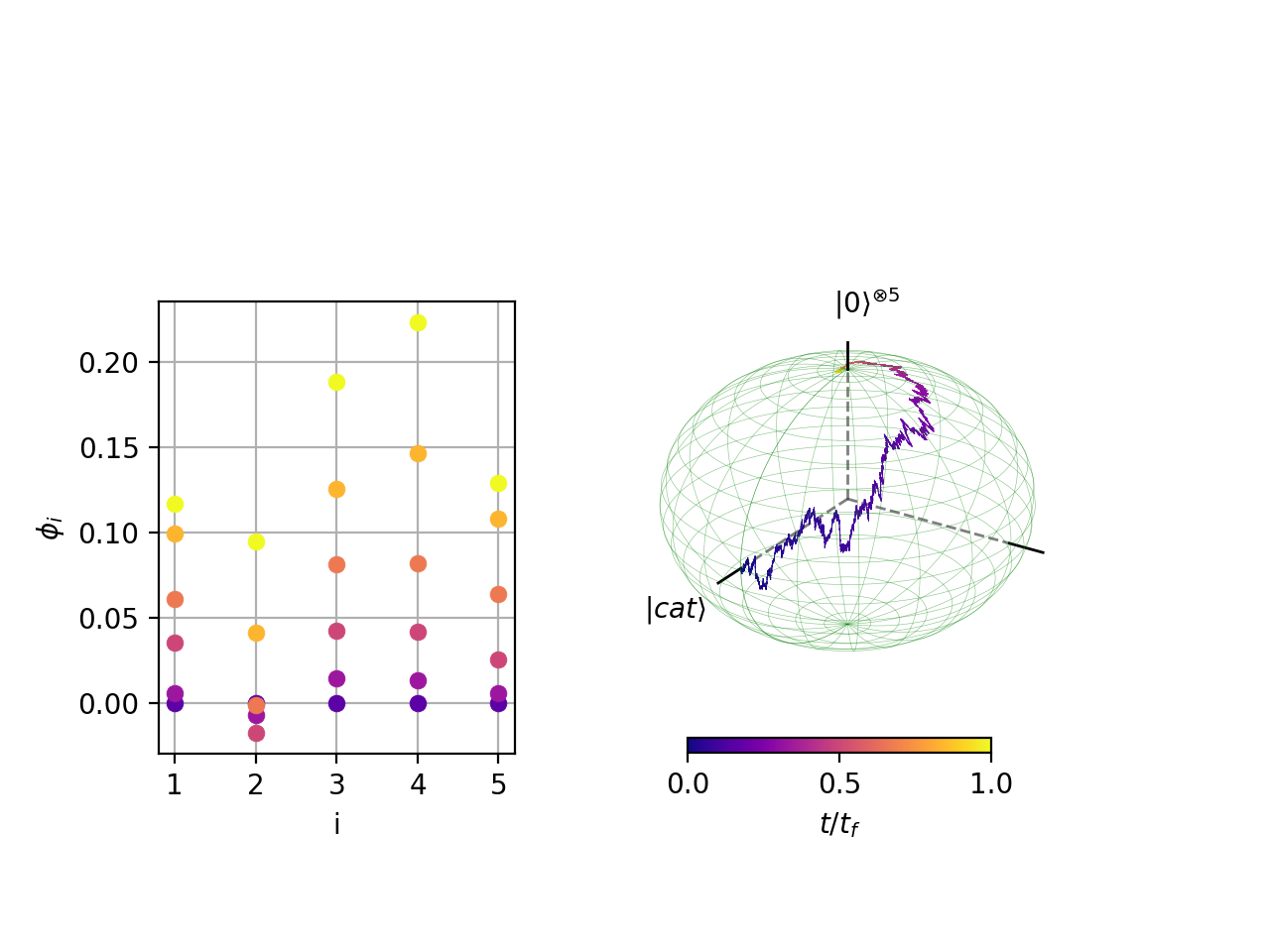}
    \caption{\footnotesize{A numerically simulated classical-quantum system of five qubits each interacting locally with a classical degree of freedom, where the $|0\rangle$ ($|1\rangle$) state acts to increase (decrease) the local classical degree of freedom $\phi_i$. Initially in a cat state, the quantum system evolves to a local product state with no entanglement, while the local classical degrees of freedom shown at six equal time intervals from $t=0$ to $t=t_f$ exhibit fluctuations around the expected increase due to diffusion. The Hamiltonian and initial conditions are provided in Appendix \ref{sec: toy}.}}
    
    \label{fig:vacuum_traj}
\end{figure}

The semi-classical dynamics we have presented also provides a framework in which to ask what happens at earlier times when there are genuine quantum fluctuations. The toy model discussed in Appendix \ref{sec: toy} and simulated in Figure \ref{fig:vacuum_traj} gives some indication of what we expect to happen. The total quantum state of the field can stay pure, even as the density of a local patch converges to a particular value. In the process, the quantum state of the field must become less entangled. The density, and thus the expansion factor at any particular point will be correlated with the density and expansion elsewhere, because the state of the field is initially highly entangled. In contrast to typical treatments, models in our framework exhibit additional fluctuations in the classical system due to diffusion, on top of the standard fluctuations due to the statistics of the decohered quantum state. Additionally, the semi-classical models presented here may be studied without a priori identifying sources of decoherence. Exploring features such as these in more realistic models would be of great interest, especially since here we can consistently evolve the system before the fluctuations have decohered.

Secondly, it is evident from equations \eqref{eq: healedSemiclass} and \eqref{eq: healedSemiQuant} that a proper treatment of semi-classical physics must take into account the fact that solutions to the dynamics should be described by ensembles of classical-quantum trajectories. In general, while the state of the quantum system conditioned on some partial classical information will be mixed, the dynamics of equation \eqref{eq: healedSemiQuant} preserves the purity of $\rho(t|\sigma\{Z_s\}_{s\leq t})$, the quantum state conditioned on the full classical trajectory, . Put more compactly, although the entropy of the quantum state conditioned on the final classical state $z_f$ may be greater than zero, the quantum state conditioned on the classical trajectory has entropy $S(\rho(t|\sigma\{Z_s\}_{s\leq t}))=0$.  If the dynamics is such that a memory record is kept of the entire trajectory, then the quantum state can also be pure conditioned on the final state of the classical system.

A central motivation for studying the semi-classical limit in general relativity is to better understand black hole evaporation, and in particular, the black hole information problem, and we briefly speculate on whether this is likely to be the case. The fact that the conditional quantum state remains pure when the decoherence vs diffusion trade-off is saturated puts an interesting twist on the black hole information problem. First, we see here that we can have a breakdown of predictability, while the quantum state remains pure. These two concepts are distinct. The purity of the quantum state is somewhat ironic, since one of the motivations 
for the present study was to produce a theory in which information could be destroyed \cite{oppenheim_post-quantum_2018}.

Secondly, we see that there are two potential categories of information loss. The first is due to treating space-time classically. Any theory in which one system is treated classically, necessarily results in diffusion of the classical system. If we believe that the present theory is fundamental, then this is a genuine source of information loss. If on the other hand, we believe that the present theory is effective, then this information loss is merely a result of approximating quantum gravitational degrees of freedom by a classical space-time. If we think of the Hawking radiation as being entangled with gravitational degrees of freedom, then treating gravity classically will destroy phase information and result in this breakdown of predictability. Viewed in this way, this form of information loss is not fundamental.

However, there is an additional category of information loss on top of this, of which there are two sources: (i) we can loose information if part of the Hilbert space becomes inaccessible to measurements; (ii) we can have information loss if we are unable to follow the full trajectory of the classical system. Both kinds of information loss could occur in spacetimes with a black hole event horizon.

It was generally believed that understanding apparent black hole information loss required a full theory of quantum gravity to describe the back-reaction onto space time. Here we see that this is not necessarily the case -- we can have information loss, even within this semi-classical framework, separate from that due to working in the framework itself. There is thus at least some potential to study the phenomena of black hole information, within the semi-classical framework presented here.


\noindent \emph{Acknowledgements.}
We would like thank Maite Arcos, Thomas Galley, Flaminia Giacomini,  Bei-Lok Hu, Claus Kiefer, Emanuele Panella, Andrea Russo, and Andy Svesko for valuable discussions. We are particularly indebted to Antoine Tilloy and Lajos Di\'{o}si for stimulating discussions on the relationship between hybrid dynamics and continuous measurement. JO is supported by an EPSRC Established Career Fellowship, and a Royal Society Wolfson Merit Award. I.L~acknowledges financial support from EPSRC. This research was supported by the National Science Foundation under Grant No. NSF PHY11-25915 and by the Simons Foundation {\it It from Qubit} Network. Research at Perimeter Institute is supported in part by the Government of Canada through the Department of Innovation, Science and Economic Development Canada and by the Province of Ontario through the Ministry of Economic Development, Job Creation and Trade.

\bibliographystyle{unsrtnat}
\bibliography{refsHealing}
\appendix
\numberwithin{equation}{section}
\newpage
\onecolumngrid

\section{Derivation of the unravelling for continuous hybrid dynamics}\label{sec: unravelingCont}

In this section, we prove that equations \eqref{eq: unravelingCQClass} and \eqref{eq: unravelingCQCQuantum} give rise to the continuous CQ master equation \eqref{eq: continuousME}. To start with, we note that by the definition $\varrho(z,t)=\mathbb{E}[\delta(Z_t-z)\rho_t]$, the dynamics of $Z_t$ and $\rho_t$ induce the following evolution on the CQ state
\begin{equation}
d\varrho(z,t)= \frac{\partial \varrho(z,t)}{\partial t}dt=\mathbb{E}[ d(\delta(Z_t-z) \rho_t) ].
\end{equation}
One must therefore calculate
 \begin{equation}\label{eq: derivativeCQ}
 \mathbb{E}[ d(\delta(Z_t-z) \rho_t) ]= \mathbb{E}[ d\delta(Z_t-z) \rho_t + \delta(Z_t-z) d \rho_t + d\delta(Z_t-z) d \rho_t ].
 \end{equation} 
 For clarity we shall go through each term individually. Using Ito's lemma with \eqref{eq: unravelingCQClass} the first term in Equation \eqref{eq: derivativeCQ} reads
\begin{equation}\label{eq: firstTerm}
\begin{split}
\mathbb{E}[ d\delta(Z_t-z) \rho_t]=& \mathbb{E}[ \frac{\partial }{\partial Z_i}[\delta(Z_t-z)]\rho_t( D_{1,i}^{00} (Z_t,t) + \langle D^{\alpha 0}_{1,i}(Z_t,t) L_{\alpha} + D^{0 \alpha}_{1,i} (Z_t,t) L_{\alpha}^{\dag} \rangle) ] dt \\
& + \mathbb{E}[\frac{1}{2}\frac{\partial^2 }{\partial Z_i \partial Z_j}[\delta(Z_t-z)] \rho_t \sigma_{ik}(Z_t,t) \sigma^T_{kj}(Z_t,t)]dt.
\end{split}
\end{equation}
To simplify Equation \eqref{eq: firstTerm} we can use some well known facts about the delta functional. Using the two identities $\partial_{Z_i} \delta(Z-z) = - \partial_{z_i} \delta(Z-z)$ and $f(Z) \delta(Z-z) = f(z) \delta(Z-z)$ for any function $f$, the right hand side of Equation \eqref{eq: firstTerm} becomes
\begin{equation}
-\frac{\partial}{\partial z_i} \mathbb{E}[\delta(Z_t-z)\rho_t( D_{1,i}^{00}(z) + \langle D^{\alpha 0}_{1,i}(z) L_{\alpha} +  D^{0 \alpha}_{1,i} (z) L_{\alpha}^{\dag} )\rangle ]dt  + \frac{1}{2}\frac{\partial^2 }{\partial z_i \partial z_j}\mathbb{E}[\delta(Z_t-z) \rho_tD_{2, ij}^{00} (z)]dt.
\end{equation}
Using the definition of the CQ state in Equation \eqref{eq: CQstateDefinition}  we arrive at 
\begin{equation}
\mathbb{E}[ d\delta(Z_t-z) \rho_t] = (-\frac{\partial}{\partial z_i}[ \cqstate(z) D_{1,i}^{00}(z)] -\frac{\partial }{\partial z_i}[\langle D^{\alpha 0 }_{1,i} L_{\alpha} \rangle \cqstate] - \frac{\partial }{\partial z_i}[ \cqstate(z) \langle D^{0 \alpha}_{1,i}  L_{\alpha}^{\dag} \rangle ]  +\frac{1}{2}\frac{\partial^2 }{\partial z_i \partial z_j}[ \cqstate(z) D_{2,ij}^{00}(z)])dt
\end{equation}
The second term in Equation \eqref{eq: derivativeCQ} is simpler to calculate and gives the pure quantum evolution terms
\begin{equation}
 \mathbb{E}[  \delta(Z_t-z) d \rho_t ] = \mathcal{L}(z)(\cqstate(z)) dt .
\end{equation}
For the final term in Equation \eqref{eq: derivativeCQ}, only the second order terms $dW^2=dt$ are relevant. Using the second positivity condition in Equation \eqref{eq:tradeoff}, i.e. that $\sigma\sigma^{-1}D_1=D_1$, then
\begin{equation}
\mathbb{E}[  d\delta(Z_t-z) d \rho_t] = \mathbb{E}[\frac{\partial }{\partial Z_i}[\delta(Z_t-z)]\rho_t D_{1,i}^{\alpha 0} (Z_t,t) (L_{\alpha} - \langle L_{\alpha} \rangle)  +  D_{1,i}^{0 \alpha } (Z_t,t) (L_{\alpha}^{\dag} - \langle L_{\alpha}^{\dag} \rangle))  ]dt,
\end{equation}
and again using the standard properties of the delta function we find 
\begin{equation}
\mathbb{E}[  d\delta(Z_t-z) d \rho_t ] = - \frac{\partial }{\partial z_i}[D_{1,i}^{\alpha 0} (z) (L_{\alpha} - \langle L_{\alpha} \rangle) \varrho(z)  + D_{1,i}^{0 \alpha } \varrho(z) (L_{\alpha}^{\dag} - \langle L_{\alpha}^{\dag} \rangle)  ]dt.
\end{equation}
Summing the three contributions  gives the equation of motion for $\partial_t \mathbb{E}[ \delta(Z_t-z) \rho_t ] = \partial_t \cqstate(z)$ to be that of the continuous master equation in Equation \eqref{eq: continuousME}.

The unravelling thus generates all possible time-continuous classical-quantum dynamics for the combined state $\cqstate(z,t)$ and is in one-to-one with the master equation \eqref{eq: continuousME}. To see that the unravelling is unique (up to statistical equivalence), suppose that there existed an alternative dynamics which yielded completely positive dynamics on $\cqstate(z,t)$. Then, since the dynamics are Markovian, the drift and variance of the classical variable is known. Given one is at the classical state $z$ in phase space at time $t$, the drift is easily calculated to be $( D_{1,i}^{00} (z) + \langle D^{\alpha 0 }_{1,i} (z) L_{\alpha} + D^{0 \alpha }_{1,i}(z) L_{\alpha}^{\dag} \rangle)$ and the variance $\sigma \sigma^T$. Since the classical trajectories are time-continuous, its evolution must therefore be generated by a stochastic differential equation with these drift and diffusion coefficients, but this is unique (up to statistical equivalence) and given by \eqref{eq: unravelingCQClass}. 

Having established uniqueness of the classical dynamics, it remains to show that the quantum dynamics given by \eqref{eq: unravelingCQCQuantum} is such that $\rho_t=\rho(t|\{Z_s\}_{s\leq t}\})$. It is worth first noting that in the case when $\sigma$ is invertible this is particularly intuitive – observations of $dZ_i$ here uniquely determine the noise processes $dW_j$, and thus Equation \eqref{eq: unravelingCQClass} may be integrated to uniquely determine the state at any later time. 

To prove this for all real-valued $\sigma$ it is convenient to first rewrite the dynamics in a vectorised form. Defining the vectors $dZ_t=(dZ_{t,1},\ldots, dZ_{t,n})^T$, $dW_t=(dW_{t,1},\ldots,dW_{t,n})^T$ for the classical stochastic processes and $L=(L_1,\ldots, L_p)^T$, $L^*=(L_1^\dag,\ldots, L_p^\dag)^T$ for the quantum Lindblad operators, the dynamics takes the form

\begin{equation} \label{eq: vecC}
dZ_t=D_1^C dt + \langle D_1^* L + D_1 L^*\rangle dt + \sigma dW_t 
\end{equation}
\begin{equation} \label{eq: vecQ}
d\rho_t = -i[H,\rho_t]dt + L^T D_0 \rho_t L^* dt -\frac{1}{2}\{L^\dag D_0^T L, \rho_t\}_+ dt + dW_t^T \sigma^{-1} D_1^* (L-\langle L \rangle) \rho_t + dW_t^T \sigma^{-1}D_1 \rho_t (L^*-\langle L^* \rangle ) .
\end{equation}
To see that the dynamics indeed satisfies $\rho_t=\rho(t|\{Z_s\}_{s\leq t}\})$, we first take the transpose of \eqref{eq: vecC} and multiply it by $(\sigma \sigma^T)^{-1}$ to find that
\begin{align}
dZ_t^T (\sigma \sigma^T)^{-1} = dW_t^T \sigma^T (\sigma \sigma^T)^{-1} + ({D_1^C} + \langle D_1^* L + D_1 L^*\rangle)^T dt \ (\sigma \sigma^T)^{-1} 
\end{align}
However, by the properties of the generalised inverse, $\sigma^{-1} = \sigma^T  (\sigma {\sigma^T} )^{-1}$ for real-valued $\sigma$. As a consequence, we find that
\begin{equation}
dW_t^T \sigma^{-1} = \big{[}dZ_t^T - ({D_1^C} + \langle D_1^* L + D_1 L^*\rangle)^T dt\big{]} (\sigma \sigma^T)^{-1}
\end{equation}
which may be inserted into equation \eqref{eq: vecQ} to give
\begin{equation}
\begin{split}
d\rho_t = &-i[H,\rho_t]dt + L^T D_0 \rho_t L^* dt -\frac{1}{2}\{L^\dag D_0^T L, \rho_t\}_+ dt \\
&+ \big{[}dZ_t^T - ({D_1^C} + \langle D_1^* L + D_1 L^*\rangle)^T dt\big{]} (\sigma \sigma^T)^{-1} D_1^* (L-\langle L \rangle) \rho_t \\ &+ \big{[}dZ_t^T - ({D_1^C} + \langle D_1^* L + D_1 L^*\rangle)^T dt\big{]} (\sigma \sigma^T)^{-1} D_1 \rho_t (L^*-\langle L^* \rangle ) .
\end{split}
\end{equation}
Since the evolution of $\rho_t$ is determined completely by that of $Z_t$, this demonstrates that indeed $\rho_t$ is unique conditioned on the classical trajectory and thus that $\rho_t=\rho(t|\{Z_s\}_{s\leq t}\}$. 


To determine the conditions for purity, we first must calculate $d \tr\{\rho_t^2\}$ for an initially pure quantum state. The Ito rules imply that
\begin{equation}
 d \tr\{\rho_t^2\} = \tr\{2\rho_t d\rho_t + d\rho_t d\rho_t\}
\end{equation} into which one may substitute \eqref{eq: vecQ}. Since the Hamiltonian and stochastic terms first order in $d\rho_t$ vanish under the trace, and only the stochastic terms are relevant at second order, we find 
\begin{equation}
\begin{split}
 d \tr\{\rho_t^2\} =& 2 \tr\{\rho_t L^T D_0 \rho_t L^* dt -\frac{1}{2} \rho_t\{L^\dag D_0^T L, \rho_t\}_+ dt\} \\
 &+ \tr\{ dW_t^T \sigma^{-1} D_1^* (L-\langle L \rangle) \rho_t\  dW_t^T \sigma^{-1} D_1^* (L-\langle L \rangle) \rho_t\} \\
 &+ \tr\{ dW_t^T \sigma^{-1} D_1^* (L-\langle L \rangle) \rho_t\  dW_t^T \sigma^{-1}D_1 \rho_t (L^*-\langle L^* \rangle ) \} \\
 &+ \tr\{ dW_t^T \sigma^{-1}D_1 \rho_t (L^*-\langle L^* \rangle )\  dW_t^T \sigma^{-1} D_1^* (L-\langle L \rangle) \rho_t\} \\
 &+ \tr\{ dW_t^T \sigma^{-1}D_1 \rho_t (L^*-\langle L^* \rangle )\  dW_t^T \sigma^{-1}D_1 \rho_t (L^*-\langle L^* \rangle ) \} .
\end{split}
\end{equation} Since for pure states $\tr \{ A\rho_t B\rho_t\}=\tr \{ A\rho_t\} \tr\{B\rho_t\}$, the terms containing $D_1$ twice or $D_1^*$ twice vanish, and the mixed terms may be rearranged into one term by taking the transpose on part of each expression and using the cyclic property of the trace. Doing so gives
\begin{equation}
\begin{split}
 d \tr\{\rho_t^2\} = &2 \tr\{\rho_t L^T D_0 \rho_t L^* dt -\frac{1}{2}\rho_t\{L^\dag D_0^T L, \rho_t\}_+ dt\} \\
 &+ 2\tr\{ \rho_t (L^\dag-\langle L^\dag \rangle) D_1^T {\sigma^T}^{-1} dW_t dW_t^T \sigma^{-1} D_1^* (L-\langle L \rangle) \rho_t\}.
\end{split}
\end{equation} Using again the relation $\tr \{ A\rho_t B\rho_t\}=\tr \{ A\rho_t\} \tr\{B\rho_t\}$ and the fact that the noise vectors satisfy $dW_t dW_t^T = \mathbb{I} dt$, the above expression reduces with some rearranging to
\begin{equation}
\begin{split}
 d \tr\{\rho_t^2\} = &2 \langle L^\dag \rangle  D_0^T \langle L \rangle \ dt  - 2\langle L^\dag D_0^T L\rangle \ dt \\
 &+ 2\langle  L^\dag D_1^T {\sigma^T}^{-1} \sigma^{-1} D_1^* L \rangle \ dt - 2 \langle L^\dag \rangle D_1^T {\sigma^T}^{-1} \sigma^{-1} D_1^* \langle L \rangle  \ dt.
\end{split}
\end{equation} To check the conditions for this to equal zero, we note that since $D_0\succeq D_1^{\dag}(\sigma \sigma^T)^{-1} D_1$ we can write $D_0^T-D_1^T {\sigma^T}^{-1} \sigma^{-1} D_1^*=B^\dag B$ and so, defining a new vector of operators $\bar{L}=BL$ rewrite the above as
\begin{equation}
\begin{split}
 d \tr\{\rho_t^2\} &= 2 \sum_{\alpha=1}^p(\langle\psi|\bar{L}_\alpha^\dag |\psi\rangle \langle\psi | \bar{L}_\alpha |\psi\rangle- \langle \psi | \bar{L}_\alpha^\dag \bar{L}_\alpha |\psi\rangle \langle \psi | \psi \rangle )dt. \\
\end{split}
\end{equation}
Since each term in the above sum is less than or equal to zero by the Cauchy-Schwartz inequality, one can check that $d \tr\{\rho_t^2\}\leq 0$ as expected. All terms are zero if and only if the $\bar{L}_\alpha |\psi\rangle \propto |\psi\rangle$, but since the $\bar{L}_\alpha$ are traceless and this must hold for all $|\psi\rangle$, it must be the case that $\bar{L}=BL=0$. Thus $B$ is zero and hence we see that $d \tr\{\rho_t^2\}=0$ for an arbitrary pure state $\rho_t$ if and only if $D_0^T=D_1^T {\sigma^T}^{-1} \sigma^{-1} D_1^*$. Since $D_0$ is Hermitian and ${\sigma^T}^{-1}\sigma^{-1}=(\sigma \sigma^T)^{-1}$ is real, taking the complex conjugate shows that the dynamics keeps quantum states pure if and only if the decoherence-diffusion trade-off \cite{decodiff2} is saturated such that $ D_0= D_1^{\dag}(\sigma \sigma^T)^{-1} D_1$. 

It is then easy to show that Equation \eqref{eq: unravelingCQPsi} is equivalent to \eqref{eq: unravelingCQCQuantum} using the standard Ito rules 
\begin{equation}\label{eq: stateVectorEvolutionApp}
    d \rho_t = d|\psi_t \rangle \langle \psi_t | + |\psi_t \rangle d \langle \psi_t | + d |\psi_t \rangle d\langle \psi_t |.
\end{equation} 

\section{The church of the larger Hilbert space and the temple of the larger phase space} \label{sec: temple}

Recall the decoherence-diffusion trade-off of Equation \eqref{eq:tradeoff},
\begin{align}
    D_0 \succeq D_1^\dag (\sigma\sigma^T)^{-1} D_1
\end{align} 
We have thus far seen that any classical-quantum dynamics of equations \eqref{eq: unravelingCQClass} and \eqref{eq: unravelingCQCQuantum} that saturate the trade-off
such that $ D_0= D_1^{\dag}(\sigma \sigma^T)^{-1} D_1$, has the property that when initially pure, both $\rho_t$ and the quantum state conditioned on the classical trajectory $\rho(t|\sigma\{Z_s\}_{s\leq t})$ remain pure. We now demonstrate that any dynamics may be purified by a dynamics that saturates the trade-off in either an enlarged quantum Hilbert space or an enlarged classical phase space. Note that this is separate from the question of whether the dynamics may be considered within an entirely quantum theory, and thus purified in a Hilbert space alone \cite{layton2024classical}.

Consider some
general dynamics given by \eqref{eq: unravelingCQClass} and \eqref{eq: unravelingCQCQuantum}. 
Defining $\tilde{D}_0= D_0-D_1^{\dag}(\sigma \sigma^T)^{-1} D_1$, then since the decoherence-diffusion trade-off is satisfied, this object must be positive semi-definite. As such, we are free to consider $D_1^{\dag}(\sigma \sigma^T)^{-1} D_1$ and $\tilde{D}_0$ to be two distinct components of the decoherence for the classical-quantum dynamics. Since the first component explicitly saturates the trade-off, the $\tilde{D}_0$ component represents the additional decoherence that prevents the quantum state being pure at all times when conditioned on the classical degrees of freedom. The idea of purifying the system will be to find some additional degrees of freedom, quantum or classical, such that when they are traced out they give rise to this additional decoherence. The evolution then  saturates the trade-off in an enlarged state space, and thus has a description in terms of pure states $|\psi\rangle_t$.

To purify the dynamics using the conventional method of an enlarged Hilbert space, we first note that the positive semi-definite matrix $\tilde{D}_0(Z_t)$ generates the following map on the quantum state at each time step $\delta t$ along a trajectory:
\begin{equation}\label{Dtildemap}
\rho_{t+\delta t}=\rho_t+ \tilde{D}^{\alpha\beta}_0(Z_t)L_{\alpha} \rho_t L_{\beta}^{\dag} \delta t - \frac{1}{2} \tilde{D}_0^{\alpha \beta}(Z_t)\{L_{\beta}^{\dag} L_{\alpha}, \rho_t \}_+ \delta t.
\end{equation}
Exploiting the singular value decomposition $\tilde{D}_0=V\Sigma V^\dag$ where $V$ is unitary and $\Sigma$ is a diagonal matrix with non-negative elements $d_\gamma$ for $\gamma=1,\ldots,\text{rank}\tilde{D_0}$, we may define the operators
\begin{equation}
M_\gamma(Z_t)=
    \begin{cases}
    \sqrt{d_\gamma }\sum_\alpha V^\alpha_\gamma L_\alpha \sqrt{\delta t} &\quad \gamma = 1,\ldots,\text{rank}\tilde{D}_0 \\
    I-\frac{1}{2}\tilde{D}_0^{\alpha\beta}L_\beta^\dag L_\alpha\delta t &\quad \gamma=\text{rank}\tilde{D}_0+1
\end{cases}
\end{equation}
 to write the map as
\begin{equation}
    \rho_{t+\delta t}=\sum_{\gamma=1}^{\text{rank}\tilde{D_0}+1} M_\gamma(Z_t) \rho_t M_\gamma^\dag(Z_t). 
\end{equation}
The map is therefore explicitly of the Kraus form and therefore CPTP. It therefore has a representation in terms of a unitary $U$ that acts on the quantum system in question and an additional environment Hilbert space $\mathcal{H}_E$ of dimension $d\geq\text{rank}\tilde{D}_0+1$. Specifically, if the unitary acts on the system and a reference state of the environment $|0\rangle$ as
\begin{equation}
U|\psi\rangle|0\rangle = \sum_{\gamma=1}^{\text{rank}\tilde{D}_0+1}M(Z_t)_\gamma |\psi\rangle |\gamma\rangle,
\end{equation} then tracing out the environment gives back the map \eqref{Dtildemap} (for more details, see for example \cite{nielsen2002quantum}). Since this is true for every $\delta t$, it must be the case that we can fully describe the evolution due to $\tilde{D}_0$ by pure states in an enlarged Hilbert space. The remaining dynamics not generated by $\tilde{D}_0$ is pure conditioned on the classical degrees of freedom, and thus the whole dynamics saturates the decoherence-diffusion trade-off in this enlarged space. Note that a more realistic purification model generated by tracing out a bath would have a Hamiltonian and therefore be explicitly of the form of Equations \eqref{eq: unravelingCQClass} and \eqref{eq: unravelingCQPsi}. However, note that in this case one would also need to make a number of approximations to regain the Lindblad form of \eqref{Dtildemap}.

It is also possible to purify the dynamics by introducing additional classical degrees of freedom, and, in contrast to the quantum case, this leads to an explicit model of purification without need for approximation. Considering again $\tilde{D}_0= D_0-D_1^{\dag}(\sigma \sigma^T)^{-1} D_1$ on the original Hilbert space (i.e. the elements $\tilde{D}^{\alpha\beta}_0$ refer to the same set of $L_\alpha$ as in \eqref{eq: unravelingCQClass} and \eqref{eq: unravelingCQCQuantum}), we now consider an enlarged phase space $\mathcal{M}\times \tilde{\mathcal{M}}$ where $\tilde{\mathcal{M}}$ has phase space degrees of freedom $\tilde{z}_i$ for $i=1,\ldots, \textrm{rank}\tilde{D}_0$. We then consider the following dynamics:
\begin{equation}
\begin{split}
   dZ_{t,i} &= ( D_{1,i}^{00} (Z_t) + \langle D^{\alpha 0 }_{1,i} (Z_t) L_{\alpha} + D^{0 \alpha }_{1,i}(Z_t) L_{\alpha}^{\dag} \rangle) dt + \sigma_{ij}(Z_t) d W_{j} 
\end{split}
\end{equation}
\begin{equation} \label{eq: SDEenlargedDOF}
\begin{split}
   d\tilde{Z}_{t,i} &=  \langle \tilde{D}^{\alpha 0 }_{1,i} (Z_t) L_{\alpha} + \tilde{D}^{0 \alpha }_{1,i}(Z_t) L_{\alpha}^{\dag} \rangle dt  + \tilde{\sigma}_{ij}(Z_t) d \tilde{W}_{j} 
\end{split}
\end{equation}
and 
\begin{equation} \label{eq: SDEenlargedQ}
\begin{split}
d \rho_t &= -i[H(Z_t), \rho_t]dt 
+ D_0^{\alpha \beta}(Z_t)L_{\alpha} \rho L_{\beta}^{\dag} dt - \frac{1}{2} D_0^{\alpha \beta}(Z_t)\{L_{\beta}^{\dag} L_{\alpha}, \rho_t \}_+ dt\\
& + D_{1,j}^{\alpha 0} \sigma^{-1}_{ji} (Z_t) (L_{\alpha} - \langle L_{\alpha} \rangle) \rho_t d W_i+  D_{1,j}^{ 0 \alpha} \sigma^{-1}_{ji}(Z_t)  \rho_t (L_{\alpha}^{\dag} - \langle L_{\alpha}^{\dag} \rangle)  d W_i\\
& + \tilde{D}_{1,j}^{\alpha 0} \tilde{\sigma}^{-1}_{ji} (Z_t) (L_{\alpha} - \langle L_{\alpha} \rangle) \rho_t d \tilde{W}_i + \tilde{D}_{1,j}^{ 0 \alpha} \tilde{\sigma}^{-1}_{ji}(Z_t)  \rho_t (L_{\alpha}^{\dag} - \langle L_{\alpha}^{\dag} \rangle)  d \tilde{W}_i, \end{split}
\end{equation}  Here, $\tilde{Z}_{t,i}$ denote the stochastic processes corresponding to degrees of freedom in $\tilde{\mathcal{M}}$, and have associated noise processes $d\tilde{W}_i$. $\tilde{D}_1$ and $\tilde{\sigma}$ can be seen in \eqref{eq: SDEenlargedDOF} to correspond to the drift and diffusion in the enlarged space, and satisfy the same requirements that $D_1$ and $\sigma$ do. Note that these are assumed to solely depend on the degrees of freedom in the original classical phase space $\mathcal{M}$. Packaging up the dynamics for $Z_{t,i}$ and $\tilde{Z}_{t,i}$ as a single classical vector over the whole phase space, these equations are of the form of Equations \eqref{eq: unravelingCQClass} and \eqref{eq: unravelingCQCQuantum} and are thus Markovian and linear on the combined CQ state.

We will then impose the condition that $\tilde{D}_1^{\dag}(\tilde{\sigma} \tilde{\sigma}^T)^{-1} \tilde{D}_1=\tilde{D}_0$. While there may be many ways of satisfying this in general, one simple and explicit construction is to consider $\tilde{D}_1=\sqrt{\tilde{D}_0}$, the principle square root of $\tilde{D}_0$. This guarantees that rank $\tilde{D}_1$ = rank $\tilde{D}_0$, and is thus a valid $\tilde{D}_1$ by the earlier choice of dimension of $\tilde{\mathcal{M}}$. Since $\sqrt{\tilde{D}_0}^\dag \sqrt{\tilde{D}_0}= \tilde{D}_0$, it suffices for $(\tilde{\sigma}\tilde{\sigma}^T)^{-1}=I$ and so we see that we can simply choose $\tilde{\sigma}=I$. It thus follows that we can always choose a suitable $\tilde{D}_1$ and $\tilde{\sigma}$ such that $\tilde{D}_1^{\dag}(\tilde{\sigma} \tilde{\sigma}^T)^{-1} \tilde{D}_1=\tilde{D}_0$.

With this condition satisfied, it is easy to check that the full decoherence-diffusion trade-off is saturated for the constructed dynamics. In particular, one has that
\begin{equation}
    D_0=\begin{pmatrix}
     D_1 \\
    \tilde{D}_1
    \end{pmatrix}^\dag
    \begin{pmatrix}
    (\sigma \sigma^T)^{-1} & 0\\
    0 & (\tilde{\sigma} \tilde{\sigma}^T)^{-1} 
    \end{pmatrix}
    \begin{pmatrix}
     D_1 \\
    \tilde{D}_1
    \end{pmatrix},
\end{equation} which is satisfied by virtue of the definition of $\tilde{D}_0$ and the above constraints on $\tilde{D}_1$ and $\tilde{\sigma}$. Since the trade-off is saturated, it follows that the dynamics of \eqref{eq: SDEenlargedQ} are purity preserving by the results of Appendix \ref{sec: unravelingCont}. Thus both $\rho_t$ and the quantum state conditioned on trajectories in the full phase space $\rho(t|\sigma\{Z_s,\tilde{Z}_s \}_{s\leq t})$ remain pure if they start pure. 

To see that the above dynamics on $\mathcal{M}\times \tilde{\mathcal{M}}$ define a purification of the starting dynamics \eqref{eq: unravelingCQClass} and \eqref{eq: unravelingCQCQuantum} on $\mathcal{M}$, consider an observer only with access to the degrees of freedom on $\mathcal{M}$. Denoting the conditioned quantum state as $\rho(t|\sigma\{Z_s\}_{s\leq t})$, at each time step one must average $\rho_t$ over the possible realisations of $\tilde{Z}_t$. Since the evolution of $\rho_t$ depends only on degrees of freedom in $\mathcal{M}$, the only information lost is the realization of the noise processes $d\tilde{W}_j$, and thus the evolution of the conditioned state $d\rho(t|\sigma\{Z_s\}_{s\leq t})=\mathbb{E}[d\rho_t|\sigma\{Z_s\}_{s\leq t}]$ is computed by averaging over these noise processes. Mathematically equivalent to the formalism in continuous measurement theory of having multiple observers or inefficient detection \cite{2006Jacob}, it follows from the rules of Ito calculus that $\mathbb{E}[\rho_t d\tilde{W}]=0$, and thus we see that the dynamics of $\rho(t|\sigma\{Z_s\}_{s\leq t})$ are exactly given by equation \eqref{eq: unravelingCQCQuantum}. It therefore follows that we recover the dynamics \eqref{eq: unravelingCQClass} and \eqref{eq: unravelingCQCQuantum} when we trace out the portion of the phase space $\tilde{\mathcal{M}}$, as originally claimed.

\section{Toy models of classical-quantum dynamics }\label{sec: toy}

In this section we discuss a few simple toy models of the general dynamics illustrated above, specifically those generated by a Hamiltonian that saturate the decoherence-diffusion trade-off. Example trajectories for each of the models are simulated using basic numerical methods and the code can be found in \cite{github2022}. The simulations include toy models of a Stern-Gerlach experiment, a spin confined in a potential, a mass in a superposition of two locations generating a gravitational potential, and vacuum fluctuations as a source for the expansion rate of the universe. 

The forms of the dynamics that we numerically study here are closely related to those used to study unravellings of quantum dynamics (c.f. Section \ref{sec: contMeasurement}), which have a long history of simulation \cite{gisin1992quantum,Wiseman_2001,breuer2002theory}. While it would be interesting to study further the convergence of the numerical methods used here by using these earlier models as benchmarks, and to find ways of improving upon them, here we will be content with simply extracting the qualitative behaviour of the toy models along single realisations of the dynamics. Comparing the number of timesteps that we use (around $N = 10^5$) to those in earlier work ($N=5000$ in \cite{gisin1992quantum}), we see that these individual trajectories are likely to be accurate, and indeed demonstrate a number of features in common with earlier work \cite{gisin1992quantum,Wiseman_2001}, such as rotation around the Bloch sphere and collapse to the $z$-axis.

\subsection{Vectorised notation}
First, we restate the healed semi-classical equations \eqref{eq: healedSemiclass} and \eqref{eq: healedSemiQuant}, this time in the vectorised notation previously used for Equations \eqref{eq: vecC} and \eqref{eq: vecQ} that is convenient for computing the dynamics in models such as these:
\begin{align} \label{eq: vecC_ham}
    dZ_{t} &=  \langle \{Z_t, H \} \rangle dt + \sigma d W_t 
\end{align}
\begin{equation} \label{eq: vecQ_ham}
\begin{split} 
     d |\psi \rangle_t &= -i H|\psi \rangle_t dt  + \frac{1}{2} dW^T_t \sigma^{-1} \big[\{Z_t, H\} - \langle \{Z_t, H\} \rangle\big] |\psi \rangle_t \\
     \quad \ \ \ \ &\ \ \ \ \ -\frac{1}{8}  \big[\{Z_t, H\} - \langle \{Z_t, H\} \rangle\big]^T(\sigma \sigma^T)^{-1}\big[\{Z_t, H\} - \langle \{Z_t, H\} \rangle\big]|\psi \rangle_t dt 
\end{split}
\end{equation}
where the sufficient and necessary condition for complete positivity is that ${(\mathbb{I}-\sigma \sigma^{-1})\{Z, H\}=a(z) \mathds{1}}$ for some $a(z)\in \mathbb{R}^n$. 

\subsection{Linear Diosi model}

In this section we turn our attention to a simple case of the Hamiltonian dynamics of Equations \eqref{eq: vecC_ham} and \eqref{eq: vecQ_ham}. In particular, we will consider a qubit coupled to a classical particle moving in one dimension. As a consequence, for any dynamics and at all times, we can characterise the classical-quantum system by a point $(q,p)$ in phase space and a point in the Bloch sphere.

We will consider dynamics generated by the Hamiltonian 
\begin{equation}
    H(q,p)=\frac{p^2}{2m}\hat{\mathds{1}}- 2 \lambda q \hat{\sigma}_z + \phi \hat{\sigma}_z,
\end{equation} corresponding to a Stern-Gerlach type interaction. The interaction couples the classical particle by a linear potential to the Pauli $\hat{\sigma}_z$ operator of the qubit, with the coupling strength determined by the parameter $\lambda\in \mathbb{R}$. Since the Hamiltonian is linear in phase-space coordinate $q$ and we use a single Lindblad operator, such CQ models which are continuous in phase space corresponds to the constant force models discussed in \cite{diosi1995quantum,diosi2011gravity,DiosiHalliwel}. Jumping models were previously simulated in \cite{oppenheim2020objective}. This same Hamiltonian constrains the qubit dynamics, which evolves according to both the interaction with the classical system, and a purely quantum Hamiltonian $\phi \hat{\sigma}_z$, for $\phi\in \mathbb{R}$. 

\begin{figure}[]
    \centering
    \includegraphics[width=10cm,trim={0cm 3cm 3cm 2.5cm},clip]{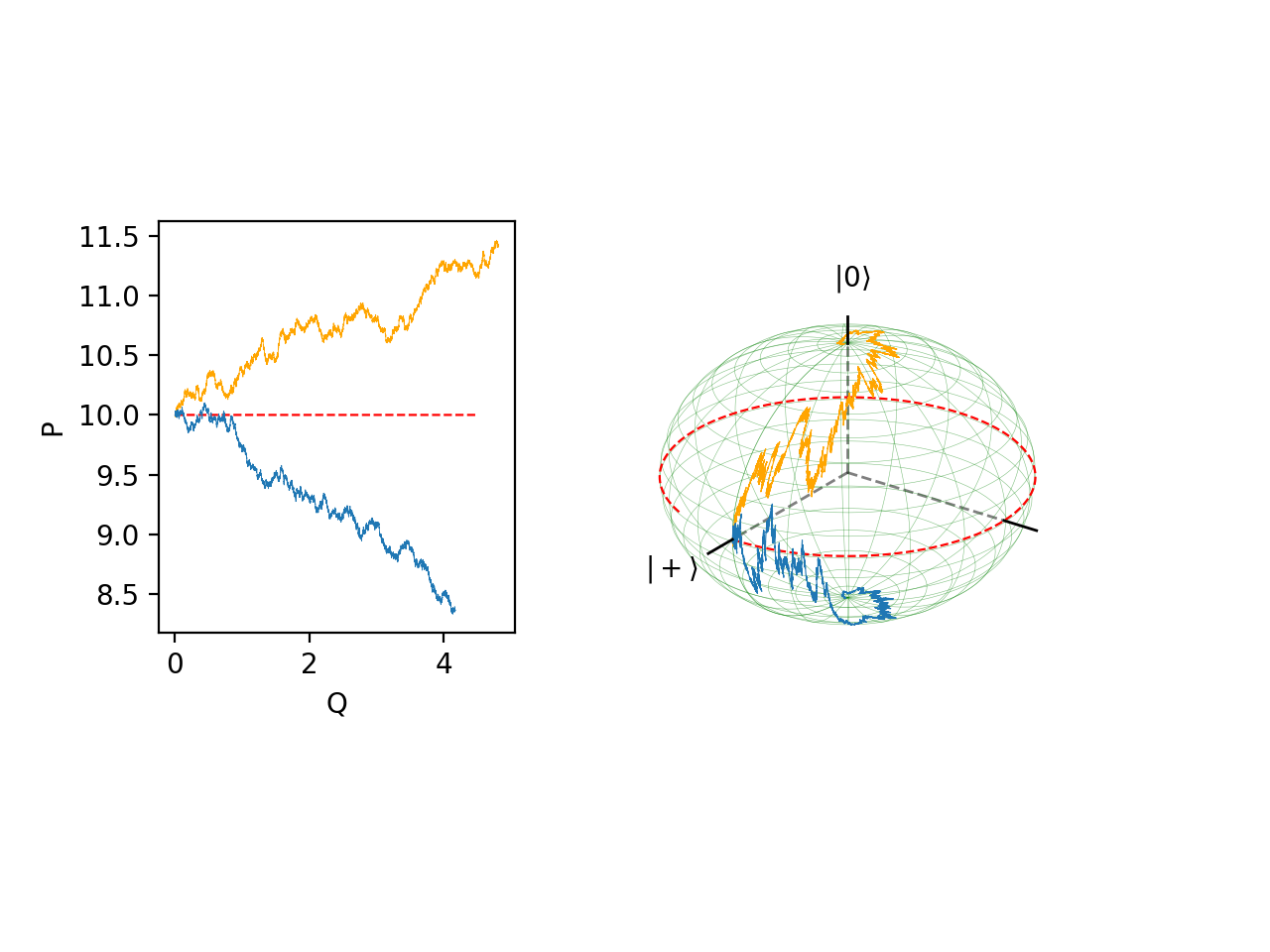}
    \caption{Classical-quantum trajectories, represented by a classical trajectory in phase space (left) and a quantum trajectory on the Bloch sphere (right), for both the standard semi-classical equations (red, dashed lines) and two distinct realisations of their healed versions (orange upper-half lines/blue lower-half lines). Here we can clearly see that in the classical and quantum trajectories are correlated, as random variables – this should be compared to the the standard semi-classical result, for which the correlations are lost. The above figure also makes it clear that if one were to average over many quantum trajectories, the fact that half move upwards and half move downwards means that there would be a loss of purity of the quantum state, with the quantum state decohering to the centre of the Bloch sphere.}
    \label{fig:this-or-this-not-this-traj} 
\end{figure}

In this case, since backreaction is only in $p$, we see that picking noise in momentum only i.e. 
\begin{equation}
\sigma=\begin{pmatrix}
    0 & 0\\
    0 & \sigma_{pp} \\
\end{pmatrix},\end{equation}
gives
\begin{equation}(\mathbb{I}-\sigma \sigma^{-1})\{z, H\} =
\begin{pmatrix}
    1 & 0\\
    0 & 0 \\
\end{pmatrix}\begin{pmatrix}
    (p/m) \hat{\mathds{1}} \\
    2\lambda\hat{\sigma}_z  \\
\end{pmatrix}= \begin{pmatrix}
    p/m \\
    0 \\
\end{pmatrix}\hat{\mathds{1}},
\end{equation} as required for complete positivity, where from now on we drop the subscript from $\sigma_{pp}\in \mathbb{R}$ for convenience. Having idenitified $H$ and a valid $\sigma$ given this, we can substitute these into Equations \eqref{eq: vecC_ham} and \eqref{eq: vecQ_ham} to find the following dynamics:
\begin{align}
dQ_t &=\frac{P_t}{m}dt \\
dP_t &= 2\lambda \langle \hat{\sigma}_z \rangle dt + \sigma dW,
\end{align}
and
\begin{align}
d|\psi\rangle_t = -i (-2\lambda Q_t + \phi)\hat{\sigma}_z|\psi\rangle dt + \frac{\lambda}{\sigma}(\hat{\sigma}_z - \langle \hat{\sigma}_z\rangle) |\psi\rangle dW - \frac{\lambda^2}{2\sigma^2}(\hat{\sigma}_z - \langle \hat{\sigma}_z\rangle)^2 |\psi\rangle dt.
\end{align}
These equations form a coupled set of stochastic differential equations, and may be easily simulated using stochastic finite difference methods such as the Euler-Maruyama or Milstein methods. An example of a classical-quantum trajectory generated by the Euler-Maruyama method for this classical-quantum dynamics is shown in Figure $\ref{fig:single_qubit_traj}$ for a classical particle initially at the origin in phase space and a quantum system initially in the state $|+\rangle$ for $m=\lambda=\sigma=1$ and $\phi=2$ between $t=0$ and $t=1$ for stepsize $\Delta t =10^{-5}$. This model is also simulated in Figure \ref{fig:this-or-this-not-this-traj} for $m=\lambda=1$, $\phi=2$, $\sigma=0.8$ and step size $\Delta t = 10^{-5}$ between $t=0$ and $t=0.45$, where dynamics given by the standard semi-classical approach, i.e. via Equations \eqref{eq: hamSemiClass} and \eqref{eq: hamSemiQuant}, is also simulated to allow a clear comparison of the two theories.

Since the current goal of simulation is only to illustrate the generic features of typical trajectories of these models, a reasonable check of the accuracy of these simulations may be made by measuring the distance of the quantum state from the surface of the Bloch sphere, which in this case has a reasonably small maximum violation of the order of $10^{-3}$. For numerical simulations leading to quantitative results, such as those requiring comparison to experiment, one would need to check the convergence in probability of the chosen numerical scheme to the random processes described by the stochastic differential equations  \cite{kloeden1992stochastic}. 

\subsection{A well and a barrier in superposition}

\begin{figure}[]
    \centering
    \includegraphics[width=10cm,trim={0 3cm 3cm 2.5cm},clip]{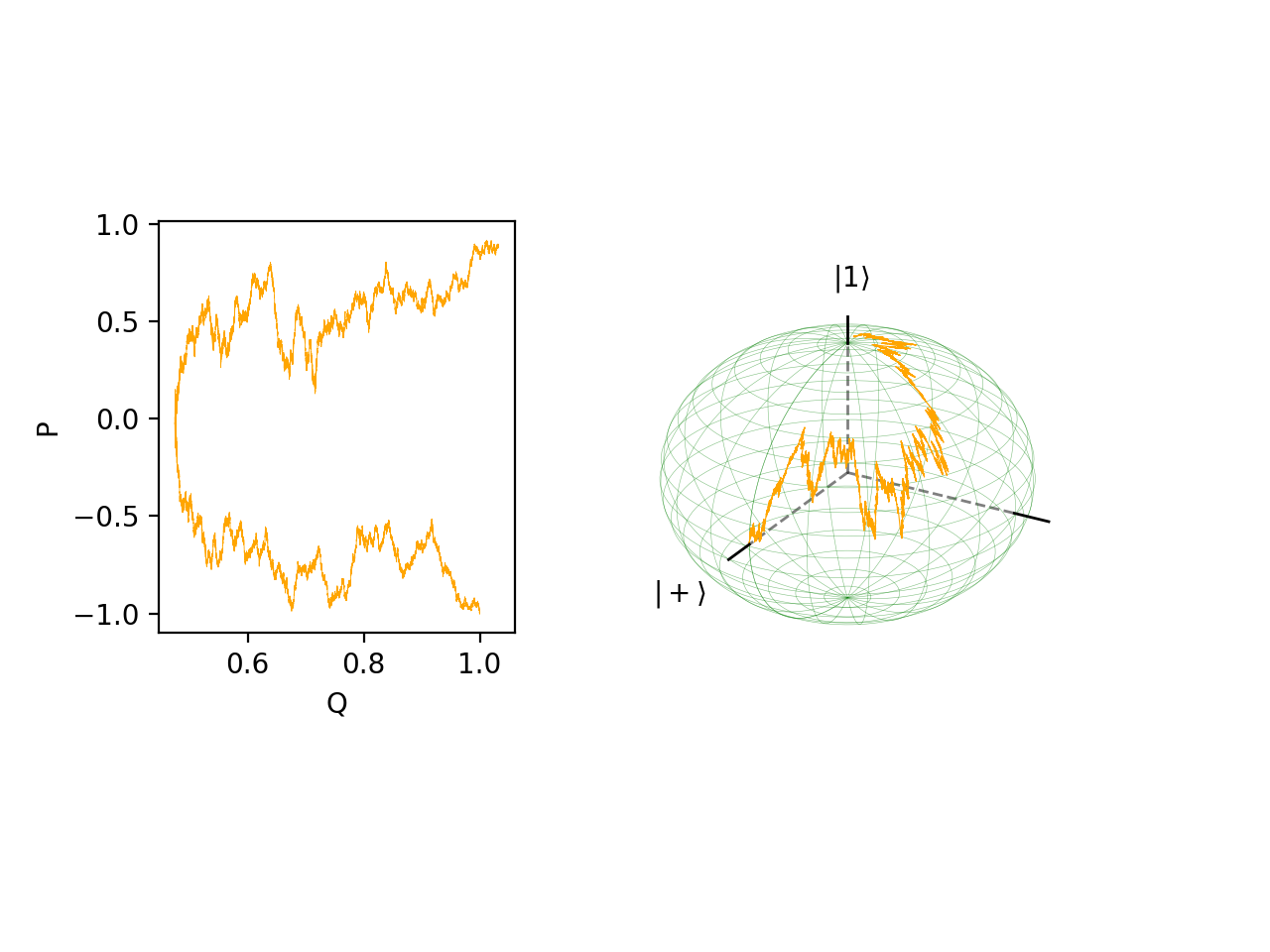}
    \caption{A classical-quantum trajectory for the $\pm\sqrt{q}$ coupling for a step size $\Delta t = 10^{-5}$ between times $t=0$ and $t=2$, and parameters $m=\lambda=1$ and $\gamma=0.5$. Initially starting at the $q=1$ and with momentum $p=-1$, the particle appears to rebound from a potential barrier $+\sqrt{q}$, agreeing with the evolution of the superpostion state $|+\rangle$ to a state close to $|1\rangle$ corresponding to a potential barrier, not a potential well. With probability $1/2$, the particle will instead encounter the potential well. At early times its evolution does not allow one to determine which of the two situations it is encountering.}
    \label{fig:root-trajectory}
\end{figure}

We now come to an example that requires the general form of CQ dynamics as presented in the main section, specifically by including an interaction Hamiltonian that is non-linear in $q$, and choosing a phase space-dependent diffusion process. Considering the qubit-particle set-up of the previous section, we now choose the Hamiltonian 
\begin{equation}
    H(q,p)=\frac{p^2}{2m}\hat{\mathbb{I}}+\lambda\sqrt{q} \hat{\sigma}_z.
\end{equation} The model describes a $\pm\lambda\sqrt{q}$ potential centred at $q=0$, i.e. either a potential well $+\lambda\sqrt{q}$ corresponding to the state $|0\rangle$ or a potential barrier $-\lambda\sqrt{q}$ for the quantum state $|1\rangle$. Although we could consider $\sqrt{|q|}$, for simplicity we will just consider the dynamics while $q>0$. To ensure the decoherence-diffusion trade-off is saturated, we will consider the form of Equations \eqref{eq: vecC_ham}/\eqref{eq: healedSemiclass} and \eqref{eq: vecQ_ham}/\eqref{eq: unravelingCQPsi}. The remaining degree of freedom is in choosing the size of the diffusion in momentum, which by the argument from the previous section is the minimal noise required, and we choose it such that $\sigma(q)=\gamma(\sqrt{q})^{-1}$ for some coupling constant $\gamma\geq0$. This gives the following dynamics
\begin{align}
dQ_t &=\frac{P_t}{m}dt \\
dP_t &= - \frac{\lambda}{2 \sqrt{Q_t}} \langle \hat{\sigma}_z \rangle dt + \frac{\gamma}{ \sqrt{Q_t}} dW,
\end{align}
and
\begin{align}
d|\psi\rangle_t = -i\lambda \sqrt{Q_t} \hat{\sigma}_z|\psi\rangle dt - \frac{\lambda} {4\gamma}(\hat{\sigma}_z - \langle \hat{\sigma}_z\rangle) |\psi\rangle dW - \frac{\lambda^2}{32\gamma^2}(\hat{\sigma}_z - \langle \hat{\sigma}_z\rangle)^2 |\psi\rangle dt.
\end{align} 
Since the strength of the noise process also increases with proximity to the centre of the potential $q=0$ by a factor of $\sqrt{q}$, the average rate of change of the quantum state is constant in time. In other words, even very close to a potential barrier, a strong repulsive force could equally be due to a large random kick in momentum by the diffusion process. As before, we simulate this model using the Euler-Maruyama method, and display the results in Figure \ref{fig:root-trajectory}.

\subsection{A toy model of a mass in superposition}
We now come to a slightly more complex example, using the same philosophy as before that the quantum state of a qubit can be used to control a potential. Here we will consider the qubit to encode the position of a mass as either on the left or the right, and consider the motion of a second test mass in the Newtonian potential generated by the heavier mass. For ease of discussion one can refer to the heavier mass as the {\it planet}, although of course the motivation comes from interest in the gravitational field of particles which can be put in superposition \cite{westphal2021measurement}. The dynamics considered here can be contrasted with that of the standard semi-classical prediction using the semi-classical Einstein equation depicted in Figure \ref{fig:notthis}. Although this is a completely consistent CQ theory, it is distinct from the models considered in Appendix \ref{sec: field}, where the gravitational field itself diffuses.

We will consider the Hamiltonian
\begin{equation}
H(\textbf{r},\textbf{p})=\frac{\textbf{p}\cdot\textbf{p}}{2m}\hat{\mathbb{I}}-\frac{GMm}{|\textbf{r}-\hat{\sigma}_z\textbf{d}|} + \phi \hat{\sigma}_z, \quad\quad \hat{\sigma}_z=|L\rangle\langle L|-|R\rangle\langle R|,
\end{equation} where $\textbf{r}$ and $\textbf{p}$ are the position and momentum of the test mass, and $\pm\textbf{d}$ is the position of the planet from the mean position. Choosing there to be only diffusion in momentum of the test particle that is given by a constant $\sigma$ for each direction, we find the dynamics for the components $i=x,y,z$ 
\begin{align}
dQ_{i} &=\frac{P_{i}}{m}dt \\
dP_{i} &=\langle\del_i{\frac{GMm}{|\textbf{r}-\hat{\sigma}_z\textbf{d}|}}\rangle + \sigma dW_i
\end{align}
and
\begin{align}
d|\psi\rangle = -i H(Q,P)|\psi\rangle dt +  \sum_{j=x,y,z} \frac{1}{2\sigma} (\del_j H(Q,P) - \langle \del_j H(Q,P) \rangle )|\psi\rangle dW_j - \sum_{j=x,y,z}\frac{1}{8\sigma^2}(\del_j H(Q,P) - \langle \del_j H(Q,P) \rangle )^2|\psi\rangle dt
\end{align} where the usual $t$ subscripts have been dropped for notational convenience, and $H(Q,P)$ and $\del_i H(Q,P)$ refer to the Hamiltonian and partial spatial derivatives of the Hamiltonian acting on the elements $Q_i,P_i$ for $i=x,y,z$.

\begin{figure}[]
    \centering
    \includegraphics[width=10cm,trim={0 3cm 3cm 2.2cm},clip]{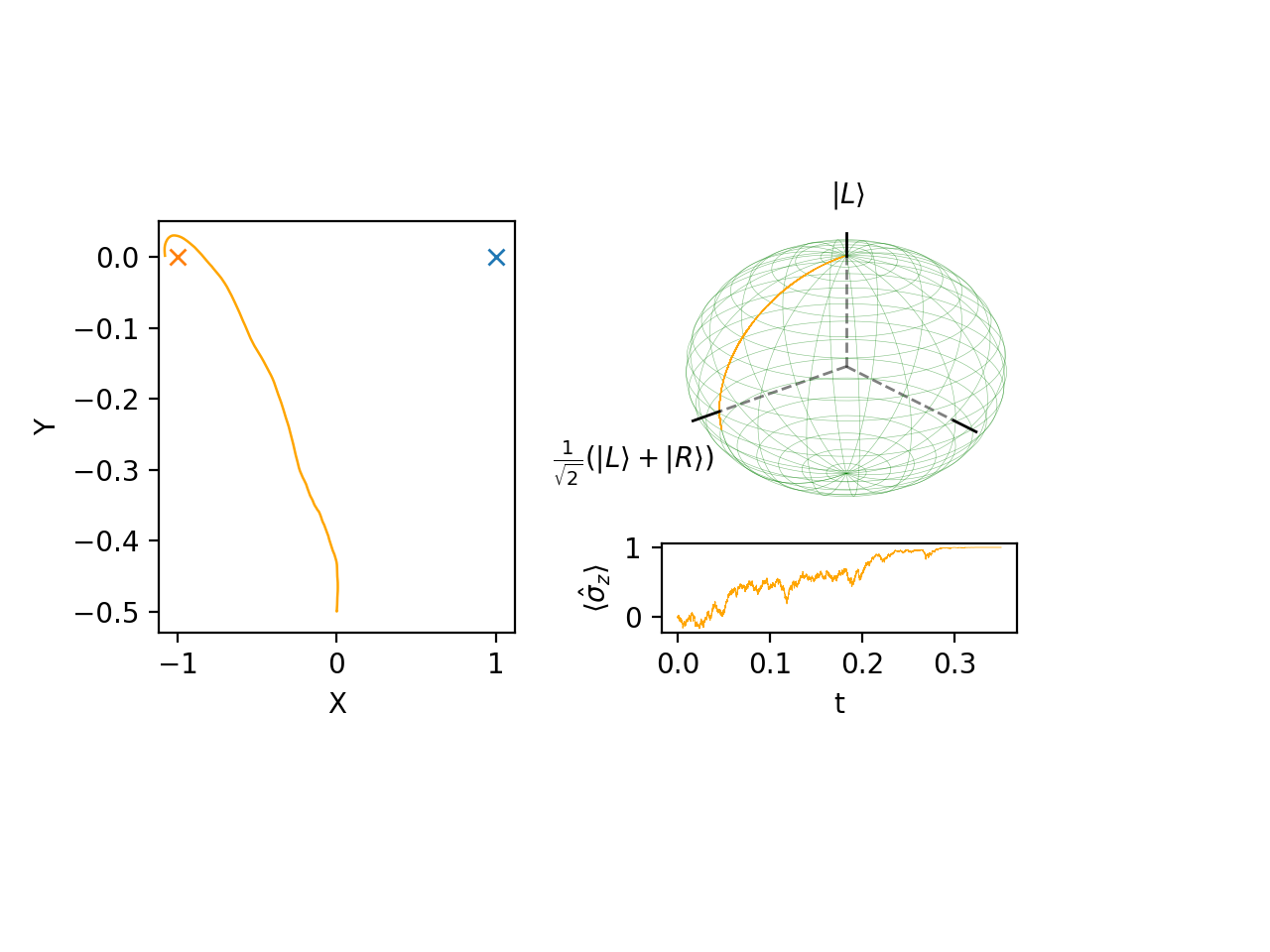}
    \caption{A classical-quantum trajectory for a test mass and a planet in superposition, for a step size $\Delta t = 10^{-5}$ between times $t=0$ and $t=0.35$, and parameters $G=1$, $M=10$, $m=0.01$, $\sigma=2m$, $\phi=0$ and $d=1$, with the momentum dimensions of phase space suppressed. Initially starting at rest at $X=0$ and $Y=-0.5$, the initial motion of the test mass towards the centre of the two masses (as predicted by the standard semi-classical theory) is due to a large random kick in momentum in the simulated realisation of the noise process. Due to negligible rotation around the z axis, the z component on the Bloch sphere is inset. }
    \label{fig:single_planet_withrot_traj}
\end{figure}

Considering first the case where $\phi=0$, an example classical-quantum trajectory is shown for this dynamics in Figure \ref{fig:single_planet_withrot_traj}, which clearly shows the test mass approaching one planet or the other. As we can see, this model gives negligible rotation around the pole of the Bloch sphere; we may arrive at a dynamics that has a clearer representation of trajectories (and equivalent physics to an observer solely monitoring the classical test particle) by letting $\phi$ be non-zero. Such a dynamics is simulated and plotted in Figure \ref{fig:single_planet_traj} for $\phi=5$, for otherwise the same choices of parameters and initial conditions as for Figure \ref{fig:single_planet_withrot_traj}.

\subsection{A toy model of vacuum fluctuations sourcing expansion}

In this toy model, we consider $n$ qubits, each coupled to a local classical degree of freedom. This provides a discretized toy model of a quantum field interacting with a classical field. We will consider the quantum system to initially be in an entangled Greenberger–Horne–Zeilinger (GHZ) or {\it cat} state \cite{greenberger1989going}, and dynamics such that each subsystem back-reacts locally on a classical degree of freedom. Since the vacuum is a highly entangled state, this simulation serves as a very crude toy model for vacuum fluctuations which source the expansion of the universe during inflation. Here, we find that the initially entangled quantum state back-reacts locally on a classical degree of freedom, so that the configurations of the classical degrees of freedom become correlated and the quantum state becomes unentangled. In the same way, we expect local expansion factors during inflation to be imprinted with correlations of the vacuum. 

Let the local classical degrees of freedom be denoted $(\phi_i,\pi_i)$ for $i=1,\ldots,n$, and the local Pauli operator be $\hat{\sigma}_z^i$. We will then take the system to evolve under the Hamiltonian
\begin{equation}
    H(\phi_1,\ldots,\phi_n,\pi_1,\ldots,\pi_n)=\lambda \sum_{i=1}^n \phi_i \hat{\sigma}_z^i + \frac{\pi^2}{2m}\hat{\mathbb{I}}.
\end{equation} Here $\lambda\in\mathbb{R}$ controls the strength of the coupling between the classical and quantum fields. As before, we will use the purity preserving Hamiltonian theory of equations \eqref{eq: vecC_ham}/\eqref{eq: healedSemiclass} and \eqref{eq: vecQ_ham}/\eqref{eq: unravelingCQPsi}, and so the only remaining freedom is in choosing the $\sigma_{ij}$. The situation here is more interesting than in the previous models, since the noise process on different lattice sites can be chosen to be correlated.  Here however we will take the simplest case and assume that the noise in momentum is  uncorrelated between lattice points with $\sigma_{p_i,p_j}=\delta_{ij}\sigma$, and that there is no diffusion in $\phi$, arriving at the dynamics
\begin{align}
d\phi_i &=\frac{\pi_i}{m}dt \\
d\pi_i &= - \lambda \langle \hat{\sigma}_z^i \rangle dt + \sigma dW_i,
\end{align}
and
\begin{align}
d|\psi\rangle = -i\lambda \sum_{i=1}^n \phi_i \hat{\sigma}_z^i |\psi\rangle dt - \frac{\lambda} {2 \sigma}\sum_{i=1}^n(\hat{\sigma}_z^i - \langle \hat{\sigma}_z^i\rangle) |\psi\rangle dW_i - \frac{\lambda^2}{8\sigma^2}\sum_{i=1}^n(\hat{\sigma}_z^i - \langle \hat{\sigma}_z^i\rangle)^2 |\psi\rangle dt,
\end{align} where the usual $t$ subscripts have been dropped for notational convenience.
While the classical degrees of freedom evolve based on the local noise and reduced quantum state, the evolution of the total n-partite quantum state is highly non-local and preserves the initial purity of the quantum state for all times. 

An example classical-quantum trajectory, where a highly entangled GHZ or {\it cat} state evolves to a local state with no entanglement, and the local classical degrees of freedom exhibit fluctuations about a mean value, is shown in Figure \ref{fig:vacuum_traj}. The simulation used a step size $\Delta t = 5\times10^{-6}$, and parameters $m=\sigma=\lambda=1$. Note that each local degree of freedom $\phi_i$ is correlated. Here the evolution of the quantum trajectory is represented by the vector $(\langle \hat{\sigma}_x^{\otimes n}\rangle,\langle \hat{\sigma}_y^{\otimes n}\rangle,\langle \hat{\sigma}_z^{\otimes n}\rangle)$, which captures the non-local dynamics that take an initial cat state $\frac{1}{\sqrt{2}}(|0\rangle^{\otimes n} + |1\rangle ^{\otimes n})$ to the $|0\rangle^{\otimes n}$ state – this follows a path on the surface of the Bloch sphere for odd $n$. Had the cat state collapsed to the $\ket{1}^{\otimes 5}$ state instead, the particle positions would be driven on average in the opposite direction.

It is important to note that the fluctuations here are entirely due to the noise process, rather than the initial state, which here provides the same force at each site. These models thus, when the low noise, $\sigma\rightarrow 0$, limit is not taken, provide additional fluctuations on top of the purely ``quantum" fluctutations due to the inital quantum state alone.

\section{Putting correlations back into the semi-classical limit}
\label{sec:superluminal}

\begin{figure}
\centering
\includegraphics[width=0.51\textwidth]{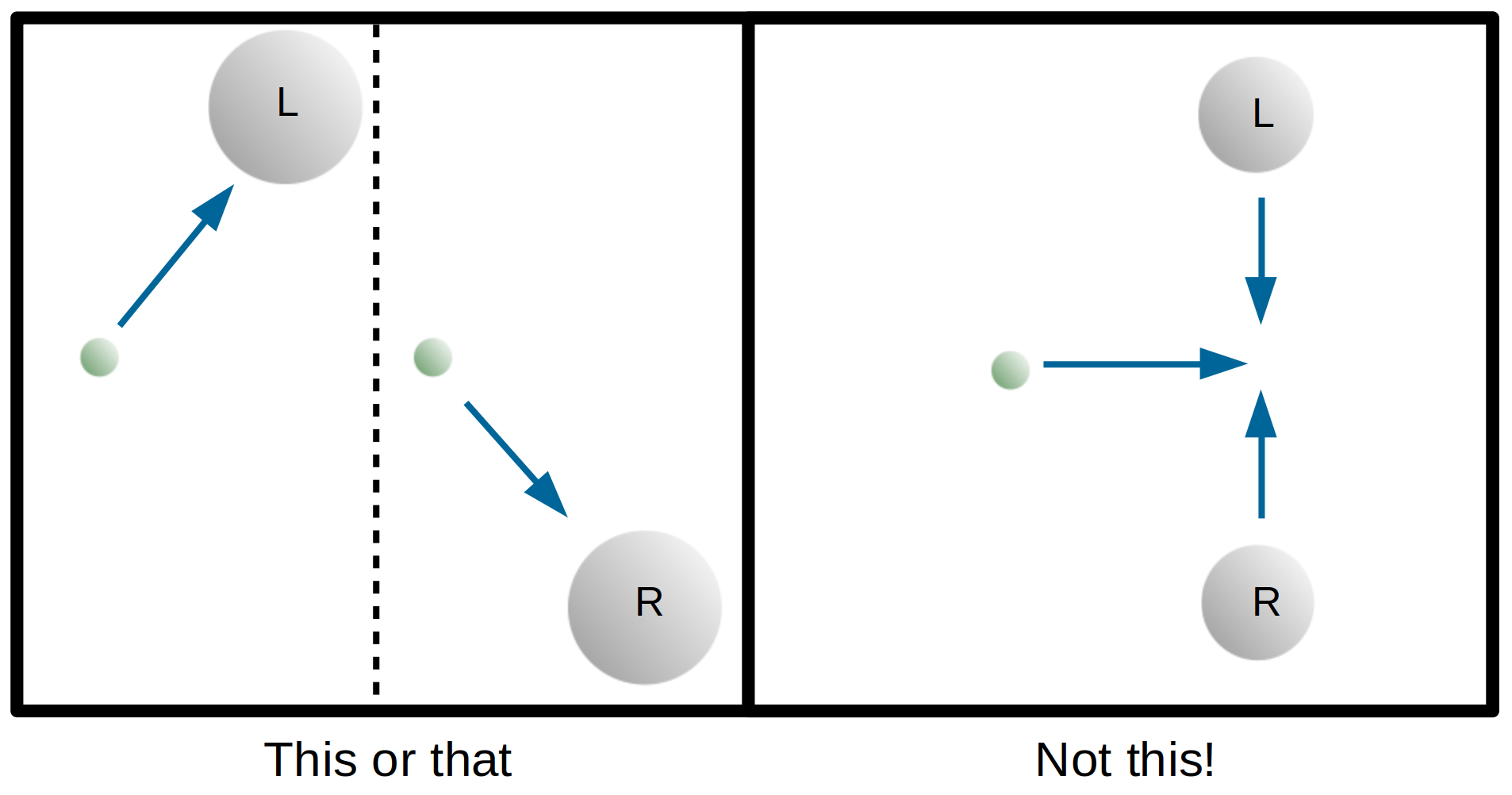}
\caption{On the left hand side, the gravitational field becomes correlated with the quantum system, a planet in a statistical mixture of being in two possible locations (``L'' and ``R''), and a test particle falls towards the planet. If one treats the semi-classical Einstein's equations as fundamental, the test particle falls down the middle which is indeed the average trajectory of its path in the left hand panel. The planets are also attracted towards the place they might have been. The culprit is that Equation \eqref{eq: semiclassEinstein} does not keep track of the correlation between the planet's position and the gravitational field.
The systems in the linear theory considered here exhibit the behaviour depicted in the left hand panel.
If the gravitational field is suitably different in the two cases, in the sense of how distinguishable the field configurations are when noise is added to them, then it may not be possible to put the planets in superposition. In the case where this is possible, the initial trajectory of the test mass will be insufficiently correlated with the planet's position to enable a determination of where the planet is. However, after the decoherence time, it may be possible to determine the planet's location from the trajectory of the test mass since its trajectory will become correlated to the planet's location. This is what one expects from a fully quantum theory, with the stochasticity of the test-mass being due to vacuum fluctuations of the metric.}
\label{fig:notthis}
\end{figure}

 It is sometimes the case that the standard semi-classical equations, Equations \eqref{eq: cControl} and \eqref{eq: semiclassEinstein} are applied in a way that is at odds with their naive interpretation. For example, in inflationary cosmology, one often considers the case where vacuum fluctuations have decohered into a statistical mixture of different matter densities. One can then solve for the expansion factor of the universe using the Friedman Robertson Walker equations for each value of the matter density. This is reminiscent of the behavior depicted in the left hand panel of Figure \ref{fig:notthis} -- the different expansion factor is correlated with different values of the matter density.  However, this is clearly not the behavior demanded by taking the expectation value of the classical statistical mixture, which would lead to a single expansion rate governed by the expected matter density. The correlation between the matter density and expansion factor, is being put in by hand, by conditioning on the matter density.

The problem is that there is no local and consistent way to perform this conditioning when you have a statistical mixture, while performing a different procedure when you have a superposition state. It violates the linearity of quantum theory in the density matrix and results in superluminal signalling \cite{Gisin:1989sx,GISIN19901,polchinski1991weinberg}. 
To see this, consider a state which has decohered into 
 a statistical mixture
of the planet being in the $\ket{L}$ or $\ket{R}$ state. It must result in something close to the behaviour depicted on the left side of Figure \ref{fig:notthis}.
Next, we prepare the maximally entangled state
\begin{align}
\ket{\psi^-}_{AB}=\frac{1}{\sqrt{2}}\left(\ket{0}_A\ket{L}_B-\ket{1}_A\ket{R}_B\right)
\nonumber
\end{align}
between a qubit held by Alice on Earth, and the distant planet that Bob orbits, where we have added a subscript $A$ ($B$) to Alice's (Bob's) ket for clarity. If Alice
measures her qubit in the $\ket{0}_A$,$\ket{1}_A$ basis, then the planet will be in a statistical mixture of being in state $\ket{L}_B,\ket{R}_B$ state, and so Bob will experience the situation depicted on the left of Figure \ref{fig:notthis}. On the other hand if Alice measures her qubit in the $\ket{\pm}_A:=\left(\ket{0}_A\pm\ket{1}_A\right)/\sqrt{2}$ basis, than one can easily check that Bob's planet will be in a statistical mixture of the $\ket{\pm}_B$. If the superposition state causes Bob to fall in a way which is different to the statistical mixture, for example falling towards the center as depicted on the right hand side of Figure \ref{fig:notthis}, then this would result in a mechanism for superluminal signalling, since Alice can choose in which basis she measures her qubit, and Bob will instantaniously know. One cannot be saved from this  conclusion by positing that Bob's trajectory when the planet is in the $\ket{-}_B$ state will be different than when it is in the $\ket{+}_B$ state, because if he can conclusively determine that Alice measured in her $\ket{\pm}_A$ basis, half the time, then that is sufficient to send signals using error correction. 

Superluminal signalling is merely a symptom of a deeper problem.  One is using a different procedure depending on the ensemble $\ket{\pm}$ versus the $\ket{L}$,$\ket{R}$ ensemble, while quantum theory demands that the density matrix alone, rather than the ensemble determine what happens. Quantum theory is linear in the density matrix. Likewise, if we have a classical probability distribution, our theory must be linear in it, for the reason outlined in Assumption \ref{ass: linear}. This means that the standard semi-classical equations are already pathological, even in the purely classical case.

On the other hand, in the master-equation approach, because it is linear and includes correlations, does ensure that the procedure is done consistently. 
Indeed, the same apparent paradox would occur in a fully quantum theory of two systems \cite{oppenheim_post-quantum_2018}.  Take for example an interaction Hamiltonian $H_{MG}=H_M\otimes H_G$, and 
let's include the correlation between the classical system and the quantum system. As in Figure \ref{fig:notthis}, when the matter system is in state $\hat\sigma^{(m)}_L$, the gravitational field will be in $\hat\sigma^{(grav)}_L$ and likewise for the $R$ states. When we have an even statistical mixture  of these two situations, the total state of both systems is
\begin{align}
\hat\sigma=\frac{1}{2}\sigma^{(m)}_L\otimes\sigma^{(grav)}_L+\frac{1}{2}\sigma^{(m)}_R\otimes\sigma^{(grav)}_R.
\end{align}
If we now let the system evolve via Heisenberg's equation of motion, and perform the trace on the quantum system, the gravitional system evolves as
\begin{align}
    \frac{\partial \sigma^{(grav)}}{\partial t}= -i\frac{1}{2}[\sigma_L^{(grav)},H_G]\tr H_M\sigma^{(m)}_L
 -i\frac{1}{2}[\sigma_R^{(grav)},H_G]\tr H_M\sigma^{(m)}_R
\end{align}
While the evolution of the system $\sigma^{(grav)}$ also appears to be determined by an expectation value, it is not
governed by the expectation value of $\tr H_M\sigma^{(m)}$ with $\sigma^{(m)}=\frac{1}{2}\sigma_L^{(m)}
+\frac{1}{2}\sigma^{(grav)}_R$, but rather, by the individual expectation value $\tr H_M\sigma_L^{(m)}$ and
$\tr H_M\sigma_R^{(m)}$.
The evolution of $\sigma^{(grav)}$ is different depending on whether $\sigma^{(m)}$ is in the $\sigma^{(m)}_L$ or $\sigma^{(m)}_R$ state, because of its correlation with the matter degrees of freedom.
This more closely resembles the situation depicted in the left panel of Figure \ref{fig:notthis}.

 Although the dynamics is linear on the combined gravitational and matter systems, it is not linear if we look at the dynamics of only the gravitational field or matter system. Since $\tr H_M\sigma^{(m)}_L\neq
\tr H_M\sigma^{(m)}_R$ the dynamics of the gravitational system is different depending on whether it is in the  $\sigma^{(grav)}_R$ or  $\sigma^{(grav)}_L$ state \cite{alicki1995comment,vstelmachovivc2001dynamics}.

\section{Unravelling by a joint classical-quantum state}\label{sec: joint}

We have seen that the most general dynamics may be unravelled by two distinct quantities, $Z_t$ and $\rho_t$, according to the coupled set of equations \ref{eq: unravelingCQClass} and \ref{eq: unravelingCQCQuantum}. However, it seems natural that we should also be able to unravel in terms of a single classical-quantum object. For instance, if we had Lindbladian dynamics on two coupled quantum systems, it would be possible to unravel the dynamics in terms of a single bipartite quantum state.

To derive such an equation, we identify the correct object as being the joint state of the system $\delta(z-Z_t)\rho_t$. The dynamics then follows almost identically to the way the CQ master equation derived from equations \ref{eq: unravelingCQClass} and \ref{eq: unravelingCQCQuantum} with Ito's lemma, the only difference being that here the expectation value is not taken. This gives the following stochastic evolution
\begin{equation} \label{eq: unravellingCQJointRho}
\begin{split}
d(\delta(z-Z_t)\rho)=& -\frac{\partial}{\partial z_i}[ \delta(z-Z_t) (D_{1,i}^{00}(z)\rho + D^{\alpha 0 }_{1, i} L_{\alpha} \rho + D^{0\alpha}_{1, i} \rho L_{\alpha}^{\dag})]dt +\frac{1}{2}\frac{\partial^2 }{\partial z_i \partial z_j}[\delta(z-Z_t)D_{2,ij}^{00}(z)\rho ]dt \\
&+\delta(z-Z_t)\mathcal{L}(z)(\rho) dt  -\frac{\partial}{\partial z_i}[\sigma_{ij}(z) \delta(z-Z_t)\rho]dW_j\\
&+ D_{1,j}^{\alpha 0} \sigma^{-1}_{ji} (z) \delta(z-Z_t)(L_{\alpha} - \langle L_{\alpha} \rangle)  \rho d W_i 
+  D_{1,j}^{ 0 \alpha} \sigma^{-1}_{ji}(z)  \delta(z-Z_t)\rho (L_{\alpha}^{\dag} - \langle L_{\alpha}^{\dag} \rangle)  d W_i  .
\end{split}
\end{equation}
As can be checked, it is the non-linearity generating the dynamics of $Z$ in equation \ref{eq: unravelingCQClass} that cancels with the non-linearity of the $\rho$ dynamics of \ref{eq: unravelingCQCQuantum}, leaving a combined dynamics that is linear up to a non-linear term proportional to the state $\delta(z-Z_t)\rho$ itself. Since we know the term is proportional to the state, and that the state is normalised, we suffer no loss of information in dropping the term, but keeping track of a now non-normalised state.

\section{Continuous classical-quantum dynamics as continuous measurement}\label{sec: contMeasurement}
Continuous measurement has previously been studied in the context of classical-quantum dynamics. Of particular relevance is the work of \cite{DiosiHalliwel, 2016Tilloy, kafri2014classical} which used continuous measurement to study hybrid classical-quantum dynamics, and used this to argue that it could be used to describe gravity beyond the standard semi-classical Einsteins Equations. The idea of these approaches is that the classical degree of freedom can sourced by the outcome of a continuous measurement result on the quantum system. Since it is being continuously measured, the quantum state then undergoes evolution according to the stochastic non-linear Schrodinger equation, and this depends on the value of the classical measurement outcome. The equation of motion for the classical degree of freedom is sourced by the outcome of the continuous measurement, which acts as a stochastic source for the classical degree of freedom, causing it to diffuse. Since the quantum system is being measured, the measurement itself leads to decoherence on the quantum system. 
In this section we show that all continuous autonomous CQ dynamics can be given such an interpretation. In particular we assume that there exists an auxiliary classical system $Z$, whose change is driven by the outcome of continuous measurements on a quantum system. However, we should emphasis that although one can simulate CQ dynamics by adding the measurement postulate in to unitary quantum theory, we believe the more appropriate implication goes the other way -- CQ dynamics allows one to describe the act of treating a measuring device or the observer as classical \cite{oppenheim_post-quantum_2018}.

We denote the trajectories of the classical systems as $Z_t$. The continuous measurement will be described by a series of POVM's described by the Kraus operators $\{\Omega_{\vec{J}}\}$ performed in the interval $[t+dt)$. The outcome of the measurement will be labeled by $J_{t,k}$, and we shall assume that this drives the auxiliary classical variable through a force term $dZ_{t,k} = D_{t,k}^{00} + J_{t,k} dt$, where we also allow for purely classical dynamics $D_{t,k}^{00}$. We could of course choose the outcome of the continuous measurement to drive the classical system in a different way, but this will be sufficient for our purposes. 

We shall consider a generalized case of \cite{Wiseman_2001} and here let the measurement $\{\Omega_{\vec{J}}\}$ at time $t$ be explicitly $Z$ dependent, i.e, we allow for the measurement to depend on the entire classical history, which we write $\{\Omega_{\vec{J}}(Z)\}$.

Specifically, consider the measurement described via
\begin{equation}
\Omega_J(Z) = 1-iH(Z_t) dt - \frac{1}{2} D_0^{\alpha \beta}(Z_t) L_{\beta}^{\dag} L_{\alpha}dt +  L_{\alpha} D_{1,i}^{ \alpha 0}(Z_t) (D_2^{-1})^{ij}(Z_z)J_{t,j}dt
\end{equation}
Then the normalization condition on the measurement 
\begin{equation}
\int d \mu_0(J) \Omega_J^{\dag} \Omega_J =1,
\end{equation}
is satisfied so long as we pick the measure $d\mu_0(J)$ to be such that 
\begin{equation}\label{eq: measure}
\int d\mu_0(J) (J_{t,i} dt) =0, \ \int d\mu_0(J) (J_{t,i} dt) (J_{t,j} dt)  = (\sigma \sigma^T)_{ij} =  D_{2,ij}dt.
\end{equation} 
and we take $D_0^{\alpha \beta} = D_{1,i}^{0 \alpha} D_{2}^{-1ij} D_{1,j}^{\beta 0}$. Equation \eqref{eq: measure} has the same statistics as a multivariate Gaussian random variable.

To calculate the mean of $J_{t,i}$ we calculate 
\begin{equation}
\int d \mu_0(J) Tr[\rho \Omega^{\dag}_J \Omega_J] J_{t,i} = \langle D_{1,i}^{0 \alpha}(Z_t) L_{\alpha}^{\dag} +  D_{1,i}^{\alpha 0 }(Z_t) L_{\alpha} \rangle + O(dt^2) ,
\end{equation}
whilst we can similarly calculate the seconds moments $J_{t,i} J_{t,j}$. These will be independent of the system $\rho$ and hence equivalent to the statistics of a Gaussian random variable with variance  $(\sigma \sigma^T)_{ij}$. As such, the statistics of the measurement outcomes can be described by the stochastic differential equation
\begin{equation}\label{eq: Jevolution}
J_{t,i} dt = dZ_t= \langle D_{1,i}^{0 \alpha}(Z_t) L_{\alpha}^{\dag} +  D_{1,i}^{\alpha 0 }(Z_t) L_{\alpha} \rangle dt + \sigma_{ij}(Z_t) W_j.
\end{equation} 
Given the outcome $J_{t,k}$. The conditioned density matrix takes the form
\begin{equation}
\rho' = \frac{\Omega_J \rho \Omega_J^{\dag}}{Tr[\Omega_J \rho \Omega_J^{\dag}]}.
\end{equation}
Then denoting $\tilde{L}^j =  D^{\alpha 0}_{1,i} (\sigma^{-1})^{ij}(Z_t) $ we find 
\begin{equation}
    \begin{split}
\rho' &= \rho - i[H(Z_t), \rho]dt - \frac{1}{2}D_0^{\alpha \beta}(Z_t) \{ L_{\beta}^{\dag} L_{\alpha} , \rho \} dt + ( \tilde{L}^i - \langle \tilde{L}^i \rangle) \rho J_{t,i} dt +  \rho ( \tilde{L}^{\dag i} - \langle \tilde{L}^{\dag i} \rangle)J_{t,i} dt \\
& + \tilde{L}^i \rho \tilde{L}^{\dag j}  J_{t,i} dt J_{t,j} dt  + D_0^{\alpha \beta}(Z_t) \langle L_{\beta}^{\dag} L_{\alpha} \rangle dt - \langle \tilde{L}^{\dag j} \tilde{L}^i \rangle \ J_{t,j} dt J_{t,i} dt + \langle \tilde{L}^i + \tilde{L}^{\dag i} \rangle  \langle \tilde{L}^j + \tilde{L}^{\dag j} \rangle J_{t,i} dt J_{t,j} dt.
\end{split}
\end{equation}
Substituting for $J_{t,i} dt$ in Equation \eqref{eq: Jevolution} one finds the continuous CQ unraveling equation which \textit{saturates} the decoherence diffusion trade-off. To obtain the general form of master equation, as in \eqref{eq: continuousME}, one can simply add extra decoherence terms the quantum state evolution. In this sense, any CQ master equation which doesn't saturate the trade-off can be interpreted as a continuous measurement process with \textit{inefficient} quantum measurements \cite{2006Jacob}. 

Within this framework, the decoherence-diffusion trade-off has a particularly simple interpretation as a manifestation of the information-disturbance trade-off. Recall that the trade-off $D_{2} \succeq D_1 D_0^{-1} D_1^{\dag}$ tells us that if we wish to minimise the impact of disturbance on one system, we must increase the disturbance on the other by an amount related to the strength of their interaction. In this case, the weaker the continuous measurement on the quantum state the smaller the coefficients in $D_0$ are, and thus smaller the resulting decoherence (with the caveats of the main text) of the quantum state. However, with weaker measurements so the resulting noise of the measurement outcomes is greater. Since these are directly input as forces onto the classical system, the greater noise in measurement outcomes leads to greater diffusion on the classical state. The stronger the coupling is, the larger the coefficient pre-multiplying the noise is, thus explaining the appearance of the coupling strength $D_1$ in the trade-off.

To see how the loss of predictability in the classical degrees of freedom in the CQ treatment may be reconciled with the complete predictability provided by a fully quantum treatment, we can utilise this alternative formulation in which the classical trajectories are generated by the measurement signals of POVMs applied to the quantum state. In this case, since every POVM may be viewed as unitary evolution on a larger system containing a measurement device, and applying unitaries controlled by the measurement device state to a set of quantum states $|z_i\rangle$ that span the classical state space, we arrive at a fully quantum, albeit artificial, model of semi-classicality. Considering the problem in discrete time, with time step $\Delta t$, the action of the unitary at a given time step $k$ is given on an arbitrary quantum state $|\psi\rangle$, the $k$th initialised measurement apparatus state $|0\rangle_k$ and the classical system as
\begin{equation}
\begin{split}
&U_k (|\psi\rangle \otimes |0\rangle_k \otimes |z_i\rangle)=
\int \Omega_J (z) |\psi\rangle \otimes |J\rangle_k \otimes |z_i+(D_{1,i}+J_i)\Delta t\rangle d\mu_0(J).
\end{split}
\end{equation} 
The preservation of the norm for all quantum states $|\psi\rangle$ follows from the definition of the POVM. While measuring the apparatus state at each time step leads to a discretized version of the CQ evolution, leaving the system unmeasured leads to a highly entangled state encoding the full probability distribution of classical-quantum trajectories in the apparatus degrees of freedom. We can equally view this in the language of many worlds, where the apparatus states $|J\rangle_k$ keep track of the branch of the wavefunction. In the case of a unique final state $z_f$ for the classical evolution, the entangled state factorises between the classical subsystem and the other two subsystems. In this case, any measurements on the final quantum state appear mixed by virtue of entanglement with an unmeasured reference system.

\section{Unraveling of classical-quantum field theory}\label{sec: field}
Since gravity is a field theory, in this section we discuss unravelings in the context of fields. The field theoretic version of the continuous master equation in \eqref{eq: continuousME} is  \cite{decodiff2}
\begin{equation}\label{eq: continuousMEFields}
\begin{split}
\frac{\partial \cqstate(z,t)}{\partial t} & =  -\int dx  \frac{\delta }{\delta z_{i}(x)  } \left( D^{00}_{1, i}(z;x) \cqstate(z,t) \right)  -\frac{1}{2}\int dx dy \frac{\delta^2 }{\delta z_{i}(x) \delta z_{j}(y) } \left( D^{00}_{2, ij}(z;x,y) \cqstate(z,t) \right)  \\
& -i[H(z), \cqstate(z,t)] + \int dx dy \left[ D_0^{\alpha \beta}(z;x,y) L_{\alpha}(x) \cqstate(z) L_{\beta}^{\dag}(y) - \frac{1}{2} D_0^{\alpha \beta}(z;x,y) \{ L_{\beta}^{\dag}(y) L_{\alpha}(x), \cqstate(z) \}_+ \right] \\
&  -\int dx \frac{\delta }{\delta z_{i}(x)} \left( D^{0\alpha}_{1, i}(z;x) \cqstate(z,t) L_{\alpha}^{\dag}(x) \right)  + \frac{\delta }{\delta z_{i}(x)} \left( D^{\alpha 0 }_{1, i}(z;x) L_{\alpha}(x) \cqstate(z,t) \right)  .
\end{split}
\end{equation}
Positivity of the dynamics is enforced by positivity of the matrix 
\begin{align} \label{eq: generalFieldMatrixInequalityMain}
\int dx dy [ b^*(x), \alpha^*(x)]  
\begin{bmatrix}
D_{2}(x,y) & D_1(x,y)\\ D_1(x,y) & D_0(x,y)
\end{bmatrix} 
\begin{bmatrix}
 b(y) \\ \alpha(y)
\end{bmatrix} \geq 0 ,
\end{align}
which should be positive for any position dependent vectors $b^{i}_{\mu}(x)$ and $a_{\alpha}(x)$ \cite{decodiff2}.
Using exactly the same methods as for the derivation of the unraveling in the case of continous classical-quantum dynamics, but replacing derivatives by functional derivatives, equation \eqref{eq: continuousMEFields} can be unraveled by the coupled stochastic differential equations
\begin{equation}\label{eq: unravelingCQField}
 \begin{split}
 & d \rho_t = \mathcal{L}(Z_t)(\rho_t)dt + \int dx dy \big[D_{1,j}^{\alpha 0}(Z_t;x) \sigma^{-1}_{ij}(Z_t;x,y) (L_{\alpha}(x) - \langle L_{\alpha}(x) \rangle) \rho_t d W_i(y)  \\
 & +  D_{1,j}^{ 0 \alpha}(Z_t;x) \sigma^{-1}_{ij}(Z_t;x,y) \rho_t(L_{\alpha}^{\dag}(x) - \langle L_{\alpha}^{\dag}(x) \rangle)  d W_i(y) \big] \\
 &  dZ_{t,i}(x) = ( D_{1,i}^{00} (Z_t;x) + \langle D^{\alpha 0 }_{1,i} (Z_t;x) L_{\alpha}(x) + D^{0 \alpha }_{1,i}(Z_t;x) L_{\alpha}^{\dag}(x) \rangle) dt + \int dy \sigma_{ij}(Z_t;x,y) d W_{j}(y) .
 \end{split}
 \end{equation}
Where now $W_{i}(x)$ is noise process satisfying
 \begin{equation}\label{eq: fieldnoise}
\mathbb{E}[W_i(x)]=0, \ \ \mathbb{E}[d W_i(x) dW_j(y)] = \delta_{ij} \delta(x,y)dt,
\end{equation}
and  $\mathcal{L}(Z)( \rho)$ shorthand for the pure Lindbladian term appearing in \eqref{eq: continuousMEFields}. In \eqref{eq: unravelingCQField} $\sigma^{-1}(x,y)$ denotes the generalized kernel inverse, so that $\int dy dz \ \sigma(x,y) \sigma^{-1}(y,z)\sigma(z,w) =\sigma(z,w) $. The equations will be local if one picks $\sigma(x,y) \sim \delta(x,y)$ but we can also allow for the more general case. In Equation \eqref{eq: fieldnoise} $\frac{d W_i(x)}{dt}$ is a white noise process, and as a result the solutions to the dynamics will in general require regularization. Studying this is beyond the scope of the current work, but a promising approach would be to study the renormalization properties of classical-quantum path integrals \cite{pathIntegralLong, covariantPI}.

\subsection{A gravitational CQ theory example}\label{sec: fullCQ}
As an example, we can use the theory of \cite{oppenheim_post-quantum_2018} to formally write down dynamics which agree with the ADM equations of motion on expectation for minimally coupled matter (we consider the Newtonian limit of this theory elsewhere \cite{UCLNewtonianLimit, decodiff2}). The idea of \cite{oppenheim_post-quantum_2018} was to assume that the dynamics is approximately Einstein gravity in the classical limit. Here we see that this fixes the drift terms, so that the Hamiltonian form of the semi-classical Einstein's equations \eqref{eq: semiclassEinstein}  are obeyed on average, and we arrive at a general form for the evolution of the 3-metric $g_{ij}$ and its conjugate momenta $\pi^{ij}$
\begin{equation}\label{eq: gravityFull}
\begin{split}
&d g_{ij} = \{ g_{ij}, H_{ADM}\lapsh \}dt,  \\
& d \pi^{ij} = \{ \pi^{ij}, H_{ADM}\lapsh \} dt - \langle \frac{\delta \qmatterham\lapsh}{\delta g_{ij}} \rangle  dt + \int dy \sigma^{ij}_{kl} [g,\pi;x,y] d W_{kl}(x) \\
& d \rho_t = -i[\qmatterham\lapsh,\rho_t] -\frac{1}{2} \int dx dy D_0^{ij;kl}[g,\pi;x,y] [\frac{\partial \qmatterham\lapsh}{\delta g_{ij}(x)}, [\frac{\partial H_m}{\delta g_{kl}(y)}, \rho_t]] dt,  \\
 & + \frac{1}{2} \int dy (\sigma^{-1})^{kl}_{ij}[g, \pi;x,y]  (  \frac{\delta \qmatterham\lapsh}{\delta g_{ij}(y)} - \langle  \frac{\delta \qmatterham\lapsh}{\delta g_{ij}(y)} \rangle ) \rho_t d W_{kl}(y) \\
 & +  \frac{1}{2}  \int dy (\sigma^{-1})^{kl}_{ij}[g, \pi;x,y] \rho_t (  \frac{\delta \qmatterham\lapsh}{\delta g_{ij}(y)} - \langle  \frac{\delta \qmatterham\lapsh}{\delta g_{ij}(y)} \rangle ) d W_{kl}(y).
\end{split}
\end{equation}
We obtain the semi-classical equation \eqref{eq: healedSemiclassField} when the dynamics is taken to be local, $\sigma\sim \delta(x,y)$, where we use equations of motion to invert $\pi_{ij}[\dot{g}]$ and obtain the expression for $G_{ij}$. Equation \eqref{eq: healedSemiclassField} is sourced by a white noise term, since $G_{ij} \sim \Ddot{g}_{ij}\sim \frac{dW_{kl}}{dt}$ which is a white noise process.  In Equation \eqref{eq: gravityFull} we can also consider the case where the lapse and shift $N,N^i$ and their conjugate momenta are included as phase space degrees of freedom. While adding them does nothing in the purely classical case when the constraints are satisfied, it does have some advantage with respect to the weak field limit \cite{UCLNewtonianLimit} and the constraint algebra \cite{Oppenheim:2020ogy} of the CQ theory. In this case, one has additional diffusion and Lindbladian terms. 
We could also add a term which describes any information loss, classical or quantum, not due to the decoherence diffusion trade-off but we have omitted such terms. Such dynamics appears to be $N$ dependent \cite{Oppenheim:2020ogy} and we make no attempt here to discuss whether or not one can make such dynamics diffeomorphism invariant leaving this as an interesting open question. Moreover, Since gravity is a constrained theory, one also expects to impose analogies of the Hamiltonian and momentum constraints in the hybrid case. A study of these was begun in \cite{Oppenheim:2020ogy}, but constraints in CQ theories are not well understood. 

In general, the dynamics will depend on the lapse and shift functions $N,N^i$ as in the Hamiltonian formulation of general relativity. On each realization of the noise process, we now have entire trajectories for each of the variables $(g_{ij}, \pi^{ij}, N, N^i)$ each associated to a quantum state, $\rho_t(t | g_{ij}, \pi^{ij}, N, N^i)$. This allows us to define a tuple $(g_{\mu \nu}, \rho_{\Sigma_t}(t))$ via the ADM embedding
\begin{equation} 
\begin{split}
\label{eq: admdecomApp}
& g_{\mu \nu} dx^{\mu} dx^{\nu}  = -N^2(t,x)dt^2 \\
&+ g_{ij}(t,x)(N^i(t,x) dt + dx^i)(N^j(t,x)dt + dx^j). 
\end{split}
\end{equation}
This associates to each trajectory a \textit{4-metric} and quantum state on a 1-parameter family of hypersurfaces $\Sigma$. The unraveling provides a method to studying dynamics of classical gravity interacting with quantum matter.

 In the gravitational context, one also expects that one should consider theories which retain diffeomorphism symmetry. Treated as an effective theory, we expect that the coefficients $D_0,D_1,D_2$ entering into the dynamics result from integrating out high energy modes of dynamical fields. As such, we should demand this  effective theory to be diffeomorphism \textit{covariant}, meaning that a solution to the dynamics in one frame should be a solution to the dynamics in any other frame, where we should transform the parameters entering into the master equation by hand, since they arise from hidden dynamical degrees of freedom.

On the other hand, if there are no degrees of freedom which have been integrated out, as would be the case in a fundamental theory of classical-quantum gravity, it is natural to impose diffeomorphism \textit{invariance} on the dynamics and the coefficients entering the master equation should be constructed out of the gravitational degrees of freedom alone. It remains to be seen if such dynamics can be made diffeomorphism invariant and give rise to full general relativity: in \cite{covariantPI} we introduce diffeomorphism invariant theories of classical-quantum gravity, which serves as a proof of principle that classical-quantum theories can be made diffeomorphism invariant, though there appears to be tension in constructing a theory which is both diffeomorphism invariant and renormalizable. Such dynamics could, in principle be taken as fundamental, and we leave it as an open question of whether or not such dynamics can be made diffeomorphism invariant with a full classical-quantum constraint surface.
\section{Markovianity and the canonical form of continuous CQ master equation}\label{sec: MarkovianApp}
In this section, we discuss how the Markovianity of the dynamics in Assumption \ref{ass: markovian} can be loosened to obtain CQ dynamics in Equation \eqref{eq: continuousME} with time-dependent parameters $D_0(t),D_1(t),D_2(t)$.

We first review various definitions pertaining to Markovianity in the purely quantum case,  before discussing the extension to the combined classical-quantum case. For a detailed discussion of Markovianity in the purely quantum context, we refer the reader to \cite{Hall_2014, Breuer_2016}
\subsection{Quantum Markovianity and its extensions}
 In quantum theory an important concept is that of a quantum operation, which is a completely positive, trace preserving (CPTP) map $\Phi_t$, $t\geq 0$, which maps density matrices $\rho$ onto themselves
\begin{equation}
    \rho_t = \Phi_t(\rho_0).
\end{equation}
Since we are interested in the dynamics of quantum systems, we shall allow the time parameter $t$ to vary $0\leq t \leq T$ which defines a one parameter family of dynamical CPTP maps $\Phi_t$.

A family of dynamics maps $\Phi_t$ is said to be \textit{time-local} if $\dot{\rho}_t = \mathcal{L}_t(\rho_t)$ for a linear map $\mathcal{L}_t$. An important feature of time-local dynamics is that they can always be written in Lindblad form \cite{Hall_2014}
\begin{equation}\label{eq: GenLindblad}
    \frac{\partial \rho_t}{\partial t}= -i[H,\rho_t] + \lambda^{\alpha \beta}(t)\left( L_{\alpha}\rho_t L_{\beta}^{\dag} - \frac{1}{2} \{L_{\beta}^{\dag} L_{\alpha}, \rho_t \} \right),
\end{equation}
where the matrix $\lambda^{\alpha \beta}(t)$ is Hermitian, but in general the conditions for complete positivity are not known \cite{Hall_2014, Breuer_2016}. In particular, though $\Phi_t$ is a completely positive map for all times $t$, $\mathcal{L}_t$ need not generate CP dynamics for $0< t \leq T$: one can consider scenarios where the initial quantum state decoheres at early times, but recoeheres at late times keeping the total dynamics $\Phi_t$ completely positive \cite{Breuer_2016}. In other words, $\mathcal{L}_t$ needs only generate CP dynamics on the subset of states  $\{ \sigma_t: \exists \ \rho_0 \text{ s.t } \sigma_t = \Phi_t(\rho_0)  \}$.

When the coefficients $\lambda^{\alpha \beta}$ are time independent and positive, then Equation \eqref{eq: GenLindblad} is the well-known Lindblad equation familiar in open quantum systems which generates CPTP dynamics \cite{lindbald, gorini_koss_sud}. We call such dynamics \textit{time-independent Markovian}, or \textit{autonomous}. The Lindblad equation represents the most general form of allowed time-local dynamics when one also demands that the generator $\mathcal{L}_t=\mathcal{L}$ be time independent \cite{lindbald, gorini_koss_sud, Pearle_2012}.

Since the generator is completely positive at $t=0$ and time independent,  $\mathcal{L}_t=\mathcal{L}$ generates completely positive dynamics on \textit{all} states \cite{lindbald, gorini_koss_sud, Pearle_2012}. This leads to an alternative, but equivalent, definition of quantum (time-independent) Markovianity which extends naturally to the time-dependent case. First, note that for Markovian dynamics one can define a two parameter family of dynamical maps as follows. Fix any two times $0 \leq s \leq t \leq T$ and consider an arbitrary initial state $\rho_s$. Then for all $0\leq s \leq t \leq T$, define the two parameter family of maps $\rho_t = \Phi_{t,s} \rho_s$ by the condition that for $s\leq t' \leq T $ the time evolution of the state obeys $\partial_{t'}\rho_{t'} = \mathcal{L}(\rho_{t'})$ for a generator $\mathcal{L}$ of Lindblad form. Conversely, consider a two parameter family of CPTP maps $\Phi_{t,s}$ such that for all  $s\leq t' \leq T $ the dynamics is time-local with a time-independent generator $\mathcal{L}$. If $\mathcal{L}$ generates CPTP dynamics on all states, then one can always write $\mathcal{L}$ in Lindblad form and the dynamics will be time-independent Markovian.

More generally one can ask that the dynamics be time-local and have $\mathcal{L}_t$ generate completely positive dynamics on all states, but relax the condition that it is time independent. In this case, the general form of master Equation is given by Equation \eqref{eq: GenLindblad} where the  coefficients $\lambda^{\alpha \beta}(t)$ are positive but are now in general time dependent. This is the quantum version of Assumption $\ref{ass: markovian}$ and has been coined \textit{time-dependent Markovianity} \cite{Wolf_2008, Tamascelli_2018, Hall_2014}. 

Though we do not use this terminology in the CQ case, it is worth mentioning that a closely related notion to time-dependent Markovianity is that of CP-divisible dynamics \cite{Breuer_2016}. A CP-divisible map can be defined when the map $\Phi_t$ is invertible. In this case, one defines a two parameter family of maps via 
\begin{equation}
    \Phi_{t,s} = \Phi_t \Phi^{-1}_s,
\end{equation}
where $\Phi_{t,0} = \Phi_t$. A family of dynamics is \textit{CP-divisble} if $\Phi_{t,s}$ is completely positive for all $0\leq s \leq t \leq T$. The connection to time-dependent Markovian dynamics arises due to the fact that when $\Phi_t$ is invertible its dynamics is always generated by a local master equation of the form in Equation \eqref{eq: GenLindblad} \cite{Anderson2007}. It can be further shown via a straightforward extension of the Lindblad equation that CP-divisible dynamics can always be written in the form of \eqref{eq: GenLindblad} with positive $\lambda^{\alpha \beta}(t)$ which can in general be time-dependent. Hence, all CP-divisible dynamics are time-dependent Markovian, but the converse is not true since time-local dynamics need not be invertible.

Note, one expects a CP map to be invertible on fairly general grounds \cite{Hall_2014}. The condition is only violated if two different states are mapped to the same state in finite time, which in practice essentially means that one reaches an equilibrium state in finite time. Since asking for time-dependent Markovianity is a weaker condition, this should also hold for a large class of physically relevant situations.
\subsection{Classical-quantum Markovianity }
Let us now extend these definitions to the combined classical-quantum case. In the CQ case, one instead starts with the notion of a classical-quantum operation $\Lambda_t$, which is a completely positive map which preserves the normalization of the CQ state $\cqstate(z)$ and the classical-quantum split \cite{poulinPC, oppenheim_post-quantum_2018}
\begin{equation}
 \cqstate(z,t)=\Lambda_t(\cqstate)(z)=\int dz' \Lambda_t(z|z')\cqstate(z',0).
\end{equation}

We call a family of classical-quantum maps $\Lambda_t(z|z')$ \textit{time-local}, if for all times
\begin{equation}
    \frac{\partial \cqstate(z,t)}{\partial t} = \int dz \mathcal{L}_t(z|z')(\rho(z',t)),
\end{equation}
for some linear CQ generator $\mathcal{L}_t(z|z')$. We call the dynamics \textit{time-independent Markovian}, or \textit{autonomous}, if the generator $\mathcal{L}_t(z|z') =\mathcal{L}(z|z')$ is time-independent. Equivalently, a CQ dynamics is time-independent Markovian when there exists a two parameter family of CQ operations  $\Lambda_{t,s}$ such that for $s \leq t' \leq t $ $\Lambda_{t,s}$ is time local and is a CP CQ operation. More generally, a CQ dynamics is \textit{time-dependent Markovian} when the generator $\mathcal{L}_t(z|z')$ is time-dependent.

When a CQ dynamics is autonomous, the generator can always be written in the form of Equation \eqref{eq: continuousME} where the coefficients $D_0,D_1,D_2$ are time independent and must satisfy the decoherence diffusion trade-off of Equation \eqref{eq:tradeoff} \cite{CQPawula}. When a CQ dynamics is time-dependent Markovian, the generator of the dynamics can still always be written in the form of \eqref{eq: continuousME} where the coefficients $D_0(t),D_1(t),D_2(t)$ are in general time dependent, but still satisfy the trade-off in Equation \eqref{eq:tradeoff} \cite{CQPawula}. However, in this later case, it is entirely equivalent to consider the coefficients to instead depend on the position $X_t$ of a free particle with trajectory $X_t=t$, which trivially changes the time-dependent Markovian dynamics  to  time-independent Markovian. Since the two kinds of Markovianity may be trivially interchanged, all of the results we present in the main body extend to the time-dependent case by adding time labels to the parameters.

If we instead asked only for the CQ dynamics to be time-local, but relaxed the Markovianity conditions for intermediate states, then - in analogy with the purely quantum case - we still expect the dynamics to still take the form of Equation \eqref{eq: continuousME}, but the decoherence-diffusion trade-off \eqref{eq:tradeoff} will be relaxed. In particular, for intermediate times we expect in the non-Markovian case one can have CP classical-quantum dynamics where the quantum degrees of freedom recohere, whilst simultaneously the classical degrees of freedom become less diffusive. It would be interesting to explore non-Markovian CQ dynamics further, since it is currently not well understood. 
\end{document}